\documentclass[useAMS,usenatbib]{mn2e}

\usepackage{psfig, epsf, epsfig}

\title[Formation of S0 galaxies]{Transformation from spirals into S0s with
bulge growth in groups of galaxies}
\author[K. Bekki and   W. J. Couch]
{Kenji Bekki${}^1$\thanks{E-mail:
bekki@cyllene.uwa.edu.au}
and Warrick J. Couch${}^2$ \\
${}^1$ICRAR M468
The University of Western Australia
35 Stirling Hwy, Crawley
Western Australia, 6009 \\
${}^2$Centre for Astrophysics and Supercomputing, Swinburne University of
Technology, Hawthorn, Victoria 3122, Australia\\}

\begin{document}

\date{Accepted, Received 2005 February 20; in original form }

\pagerange{\pageref{firstpage}--\pageref{lastpage}} \pubyear{2005}

\maketitle

\label{firstpage}

\begin{abstract}

Recent observations have revealed that the time evolution
of the S0 number fraction at intermediate and high redshifts ($0.2<z<0.8$) 
is more dramatic in groups of galaxies than in clusters. 
In order to understand the origin of S0s in groups,
we investigate numerically the  
morphological transformation of spirals
into S0s through group-related physical processes. 
Our chemodynamical  simulations show that
spirals in group environments can be strongly influenced
by repetitive slow encounters with
group member galaxies
so that those with thin disks
and prominent spiral arm structures
can be transformed into S0s  with thick disks 
and without prominent spiral arm structure.
Such  tidal interactions 
can also trigger repetitive starbursts within the bulges of spirals
and consequently increase significantly the masses of their bulges.
Owing to rapid consumption of gas initially in spirals during the bulge growth,
the S0s can become gas-poor. 
The S0s transformed from spirals in this way 
have young and metal-rich stellar populations
in the inner regions of their  bulges.
The simulated S0s have lower maximum rotational velocities and flatter radial
line-of-sight velocity dispersion profiles in comparison to their progenitor
spirals.
The formation processes of S0s due to tidal interactions
depend not only on the masses and orbits of the progenitor spirals, but also on
group mass. 
A significant fraction ($10-30$\%) of stars and gas can be stripped
during this spiral to S0 morphological transformation so that
intragroup stars and gas can be formed.
Based on these results, we discuss structures, kinematics,  chemical properties,
and the Tully-Fisher relation 
of S0s in groups.

\end{abstract}


\section{Introduction}

Many authors have long discussed when and how S0s were formed in
different environments based on their observed fundamental properties 
such as apparent shapes  (e.g., van den Bergh 1976; Holden et al. 2009), 
gas content (e.g., Bothun 1982; Welch \& Sage 2003),
stellar structure and kinematics (e.g., Kormendy \& Illingworth 1982;
Burstein 1979; Fisher 1997; Emsellem et al. 2007; Laurikainen et al. 2010),  
counter-rotating components (e.g., Bertola et al. 1992),
stellar population ages (e.g., Kuntschner \& Davies 1998),
color-magnitude relation (e.g., Ellis et al. 1997;
Stanford et al. 1998; van Dokkum et al. 1998),
and the morphology-density relation 
(e.g.,  Dressler 1980).
Most of these studies considered that S0s originate from spirals
and thereby discussed what physical roles galaxy environments (e.g., groups
and clusters) play in transforming spirals into S0s. 
Recent observations have suggested that not only environment but also
galaxy mass might be important in S0 formation 
(e.g.,  Tasca et al. 2009; Vulcani et al. 2010).
The origin of S0s has also been linked observationally 
to the evolution of other galaxy populations such
as dusty starburst galaxies (e.g., Geach et al. 2009)
and poststarburst ones  with ``E+A'' spectra (e.g., Pracy et al. 2009).

A growing number of observational studies have
revealed a smaller fraction of S0 galaxies in distant
clusters of galaxies (relative to that observed in present day clusters), 
suggesting dramatic morphological evolution
of spirals into S0s within these environments
(Dressler et al. 1997; Couch et al. 1998; van Dokkum et al. 1998;
Fasano et al. 2000; Smith et al. 2005; Postman et al. 2005;
Desai et al. 2007; Poggianti et al. 2008; Vulcani et al. 2010),
though such S0 evolution
may not be real but due to morphological classification errors (e.g., Andreon 1998).
Recent photometric studies of the structure of S0s 
have revealed intriguing scaling-relations between their bulges and disks 
and differences in bar strength between spirals and S0s 
(e.g., Buta et  al. 2010; Laurikainen et al. 2010). 
These observations have provided  new clues as to whether and how
S0s are formed from spirals in
different environments.

A number of theoretical models for the  
transformation of spirals into S0s have been proposed, which includes
ram pressure stripping (e.g., Gunn \& Gott 1972; Farouki \& Shapiro 1980;
Abadi et al. 1999), tidal encounters (e.g., Icke 1985), tidal compression
by the gravitational field of clusters (e.g., Byrd \& Valtonen 1990),
truncation of gas replenishment (e.g., Larson et al. 1980; Bekki et al. 2002),
and minor and unequal-mass merging (e.g., Bekki 1998).
Given that S0s exist in different environments such as the field, groups, and clusters, 
with the S0 fraction depending on local galaxy density (e.g., Dressler 1980),
all these physical mechanisms could be important 
and their relative importance of would depend on environmental
parameters such as  the total group and cluster mass.
Thus it remains unclear which theoretical model is the most important 
in S0 formation for a given environment.

Recent observational studies on the evolution of the S0 fraction in groups
and clusters have found that S0 evolution is significantly more
dramatic in groups than  in clusters (e.g., Wilman et al. 2009; Just et al. 2010).
These observations suggest that cluster-related physical processes such as
ram pressure stripping (e.g., Gunn \& Gott 1972) are not responsible for
the formation of the majority  of S0s.
Also, it is well known that galaxy bulges in S0s are systematically more 
luminous than those in spirals (e.g., Dressler 1980; Simien \& de Vaucouleurs 1986;
Christlein \& Zabludoff 2004), which implies that bulges in spirals need to
grow significantly in order to be transformed into S0s: disk-fading alone, due
to the truncation of star formation, cannot be the main mechanism of S0 formation.
Thus galaxy interactions and merging, which are highly likely to occur in groups,
are a promising mechanism for S0 formation and thus need to be investigated 
theoretically using numerical simulations.

The purpose of this paper is thus to investigate how
S0s are formed in the group environment using chemodynamical
numerical simulations that enable us to investigate both dynamical
properties (e.g., stellar kinematics) and star formation and chemical
evolution histories for spirals within groups in a fully
self-consistent manner. 
We consider that both (i)\,slow tidal encounters of galaxies
with relative velocities ($V_{\rm r}$) less
than $400-500$ km s$^{-1}$, and (ii)\,group tides 
are responsible for the transformation of spirals into
S0s in the group environment. We particularly investigate how spirals 
with smaller bulges are transformed
into S0s with bigger ones as a result of the tidal fields within groups and
group member galaxy interactions. We discuss how this transformation process
depends on the masses and orbits of galaxies, the interaction histories of 
group members, and the physical properties of groups such as their masses and 
sizes, based on a large and systematic parameter study of S0 formation.

The plan of the paper is as follows: In the next section 
we describe our numerical model for S0 formation  in groups. 
In \S 3, we
present the numerical results
mainly on the physical properties of the simulated S0s.
In \S 4, we discuss these results in the context of
key observational results on the physical properties of S0s.
We summarize our  conclusions in \S 5.

\begin{table*}
\centering
\begin{minipage}{175mm}
\caption{Description of the fiducial model and the range of model parameters for
other representative models.}
\begin{tabular}{cccccccccc}
{Model}
& {$M_{\rm d}$
\footnote{The initial disk mass of a spiral  in units of ${\rm M}_{\odot}$.}}
& {$R_{\rm d}$
\footnote{The initial size of a stellar disk for a spiral  in units of kpc.}}
& {$f_{\rm b}$
\footnote{The initial bulge mass fraction ($M_{\rm b}/M_{\rm d}$)
of a spiral.}}
& {$f_{\rm g}$
\footnote{The initial gas mass fraction ($M_{\rm g}/M_{\rm d}$)
of a spiral.}} 
& {$M_{\rm gr}$
\footnote{The initial total mass of  a group of galaxies 
in units of  $M_{\odot}$. }} 
& {Orbital types
\footnote{``Normal model''  means that a spiral can orbit the center
of its host group whereas ``Infall model''
means that a spiral can infall onto a group from the outside of the group. }}
& { $R_{\rm ini}$
\footnote{The initial distance of a spiral from the center of its host group.
This $R_{\rm ini}$ is the $x$-position of the spiral (and always positive)
and $r_{\rm s}$
represents the scale-length of the adopted NFW dark matter halo of the group.}}
& { $f_{\rm v}$
\footnote{For a normal model, the initial $y$-component velocity of a spiral is given as
$f_{\rm v} V_{\rm c}$, where $V_{\rm c}$ is the circular velocity
at the position of the spiral in a group.
For an infall model, the $x$-component velocity of a spiral
is given as $-f_{\rm v} V_{\rm c}$ (because $R_{\rm ini}$ is always positive
in the present study).}}
& {$b_{\rm y}$
\footnote{The impact parameter of the orbit of a spiral
in a infall model. This $b_{\rm y}$ corresponds to the initial $y$-component
of the location of the spiral.}}  \\
Fiducial & $6 \times 10^{10} $  & 17.5 & 0.17 & 0.1 
& $2 \times 10^{13}$  & Normal  &  $4r_{\rm s}$  & 0.5 & $-$  \\
Representative & (1.2-60)$\times 10^{9} $  & 6.6-17.5 & 0-0.7 & 0-0.5
& (0.05-1.0)$\times 10^{14}$  & Normal/Infall   &  (1-7)$\times r_{\rm s}$  
& 0.2-1 & (0.5-2)$\times r_{\rm s}$  \\
\end{tabular}
\end{minipage}
\end{table*}

\section{The model}

\subsection{Group of galaxies}

In order to simulate the time evolution of spirals
in groups, 
we use the latest version of GRAPE
(GRavity PipE, GRAPE-7), which is the special-purpose
computer for gravitational dynamics (Sugimoto et al. 1990).
We use our original GRAPE-SPH code (Bekki 2009)
which combines
the method of smoothed particle
hydrodynamics (SPH) with GRAPE for calculations of three-dimensional
self-gravitating fluids in astrophysics.
In the present models,  both  a spiral galaxy
and the dark matter halo of its host group of galaxies 
are represented by N-body particles so that
not only dynamical friction of the spiral against the dark matter halo of the group
but also dynamical influences of the halo on the spiral can be self-consistently
investigated. Group member galaxies other than the spiral
are modeled as point-mass particles, because we consider that
such modeling is enough to grasp the essential ingredients of tidal influences
of group member galaxies on the evolution of the spiral in the present study.

The structure of a group  is modeled
using an ``NFW'' profile predicted by
the cold dark matter cosmology (Navarro et al. 1996) as follows:
\begin{equation}
{\rho}(r)=\frac{\rho_{0}}{(r/r_{\rm s})(1+r/r_{\rm s})^2},
\end{equation}
where  $r$, $\rho_{0}$, and $r_{\rm s}$ are
the spherical radius,  the characteristic  density of a dark halo,  and the
scale
length of the halo, respectively.
The mass and size of a group  are represented by
$M_{\rm gr}$ and $R_{\rm gr}$, respectively.
The $c$ parameter ($=r_{\rm s}/r_{\rm vir}$,
where  $r_{\rm vir}$
is the  virial radius  of the NFW profile)  for a group 
with $M_{\rm gr}$
is chosen according to the predicted $c$-$M_{\rm gr}$ relation
in the $\Lambda$CDM simulations (e.g., Neto et al. 2007) :here 
the dark matter mass of the group corresponds to $M_{\rm gr}$.
A reasonable value of $c$ is thus 
5.6 for
$M_{\rm gr}=2 \times 10^{13} {\rm M}_{\odot}$ and
4.7 for
$M_{\rm gr}=10^{14} {\rm M}_{\odot}$.
We mainly show the results of  the models with 
$M_{\rm gr}=2 \times 10^{13} {\rm M}_{\odot}$
and $R_{\rm gr}=535$ kpc,  though we investigate
different models with $M_{\rm gr}$ ranging from 
$5 \times 10^{12} {\rm M}_{\odot}$
to $10^{14} {\rm M}_{\odot}$.

Galaxies in a group are represented by collisionless particles
and their spatial distribution follows the NFW profile
with $c=3$,
 which is consistent with recent $K$-band observational
studies on galaxies distributions in groups and clusters  (e.g., Lin et al.  2004).
The canonical Schechter function
(with slope of $-1.07$)
is adopted for generating a galaxy luminosity/mass function for
luminosities ranging from $0.01L^{\ast}$ to $2.5L^{\ast}$
in a group.
We assume that the mass-to-light-ratio (here mass includes dark matter) for
each individual galaxy is 20 and thereby calculate the total mass according
to the allocated luminosity.
The total mass (thus number)  of galaxies in a group 
with $M_{\rm gr}$ is determined by the mass-to-light-ratio,
that itself is dependent on $M_{\rm gr}$
(Marinoni \& Hudson  2002); 
\begin{equation}
M_{\rm gr}/L_{\rm gr}=350 {(\frac{M_{\rm gr}}{5\times 10^{14} {\rm
M}_{\odot}})}^{0.335},
\end{equation}
where $L_{\rm gr}$ is the total luminosity of galaxies in the group.
Therefore, firstly
$L_{\rm gr}$ for a group with mass $M_{\rm gr}$ is derived from the above 
equation (2) and then
the total number of galaxies for the group is determined from 
the derived $L_{\rm gr}$ and the adopted luminosity function of galaxies. 
As an example,
a group with $M_{\rm gr}= 2 \times 10^{13} {\rm M}_{\odot}$
has 87 galaxies.
Group  member  galaxies in a group  have an isotropic
velocity dispersion just as the  dark matter of the cluster does.

\subsection{Spiral galaxy}

The total  mass and the size of the disk of a spiral galaxy
with total galaxy mass $M_{\rm gal}$ (inclusive of dark matter)
are $M_{\rm d}$ and $R_{\rm d}$, respectively.
Henceforth, all masses and lengths are measured in units
of $M_{\rm d}$ and $R_{\rm d}$, respectively, unless
specified. Velocity and time are measured in units of $v$ = $
(GM_{\rm d}/R_{\rm d})^{1/2}$ and $t_{\rm dyn}$ = $(R_{\rm
d}^{3}/GM_{\rm d})^{1/2}$, respectively, where $G$ is the
gravitational constant and assumed to be 1.0 in the present
study. If we adopt $M_{\rm d}$ = 6.0 $\times$ $10^{10}$ $ \rm
M_{\odot}$ and $R_{\rm d}$ = 17.5 kpc as a fiducial value, then
$v$ = 1.21 $\times$ $10^{2}$ km/s and $t_{\rm dyn}$ = 1.41
$\times$ $10^{8}$ yr, respectively.
The disk is composed of a dark matter halo,
a stellar disk, a stellar bulge, and  a gaseous disk.
Gaseous halos that were included in our previous works (e.g., Bekki 2009)
are not included in the present paper, because we do not discuss
ram pressure stripping of halo gas within groups.

The mass ratio of the dark matter halo to the stellar disk
in a spiral  model is
fixed at 9 for most models and
the density distribution of the halo is represented by the NFW profile.
We consider that a reasonable value of the $c$-parameter 
is 7.8 (consistent with the results by Neto et al. 2007) for
Milky Way-type disk galaxies with $M_{\rm d} \approx 6 \times 10^{10} {\rm M}_{\odot}$.
The value of $r_{\rm s}$ is chosen such
that
the rotation curve of a disk is reasonably consistent with observations:
the maximum circular velocity of a Milky Way-like  disk galaxy in the fiducial model 
is 245 km s$^{-1}$.

The bulge of a spiral has a mass of $M_{\rm b}$, a size of $R_{\rm b}$
and a scale-length of $R_{\rm b,0}$
and is represented by the Hernquist
density profile. The bulge is assumed to have isotropic velocity dispersion
and the radial velocity dispersion is given according to the Jeans equation
for a spherical system.
The bulge-mass fraction ($f_{\rm b}=M_{\rm b}/M_{\rm d}$) is 
a free parameter. 
We mainly investigate ``Milky Way'' models (referred to as ``MW'' 
from now on) in which $f_{\rm b}=0.167$ and $R_{\rm b}=0.2R_{\rm d}$
(i.e.,  $R_{\rm b,0}=0.04R_{\rm d}$).
In order to determine $R_{\rm b}$ for a given $M_{\rm b}$ and $f_{\rm b}$,
we use the Faber-Jackson relation ($L \propto {\sigma}^4$,
where $L$ is the total  luminosity of a galaxy  and 
$\sigma$ is the central velocity dispersion;
Faber \& Jackson 1976) and the virial theorem.
The ``B/T ratio'' is defined as $B/T=M_{\rm b}/(M_{\rm d}+M_{\rm b})$,
which means that it is different from $f_{\rm b}$. The initial and final
$B/T$ are referred to as $(B/T)_{\rm i}$ and $(B/T)_{\rm f}$, respectively.

The radial ($R$) and vertical ($Z$) density profiles of the stellar disk are
assumed to be proportional to $\exp (-R/R_{0}) $ with scale
length $R_{0}$ = 0.2 and to ${\rm sech}^2 (Z/Z_{0})$ with scale
length $Z_{0}$ = 0.04 in our units, respectively.
In addition to the
rotational velocity caused by the gravitational field of disk,
bulge, and dark halo components, the initial radial and azimuthal
velocity dispersions are assigned to the disc component according to
the epicyclic theory with Toomre's parameter $Q$ = 1.5.  The
vertical velocity dispersion at a given radius is set to be 0.5
times as large as the radial velocity dispersion at that point,
as is consistent with the observed trend of the Milky Way (e.g.,
Wielen 1977).

We investigate  models with different $M_{\rm gal}$ and adopt
Freeman's law (Freeman 1970) to determine $R_0$
of a disk galaxy according to its disk mass:
\begin{equation}
R_{\rm 0}=C_{\rm d} {(\frac{M_{\rm d}}{6\times 10^{10} {\rm
M}_{\odot}})}^{0.5} {\rm kpc,}
\end{equation}
where $C_{\rm d}$ is a normalization constant.
We consider that $C_{\rm d}=3.5$ kpc is a reasonable value,
because it is consistent with the disk structure of the Galaxy.
We thus adopt $C_{\rm d}=3.5$ kpc as a standard value for
luminous disk galaxies.  However, as shown in Kauffmann et al. (2003),
low-luminosity disk galaxies have low surface stellar densities 
and thus $R_{\rm 0}$ determined by the above equation  would not be
so appropriate (i.e., significantly smaller than the observed).
We thus adopt $C_{\rm d}=6.2$\,kpc and 9.3\,kpc to model low-luminosity
and low-surface-brightness galaxies.
The galaxies with $C_{\rm d}=3.5$ kpc
are referred to as ``high-surface-brightness galaxies (HSBs)'' 
and those with $C_{\rm d}=9.3$\,kpc are as 
``low-surface-brightness galaxies (LSBs)''
just for convenience.
For example LSBs with $M_{\rm d}=1.2 \times 10^9 {\rm M}_{\odot}$
have $R_0=1.3$ kpc for $C_{\rm d}=9.3$ kpc.
Structural and kinematical properties of dark matter halos
and stellar disks are assumed to be self-similar between
models with different $M_{\rm gal}$.

\subsection{Gas, star formation, and chemical evolution}

Following our previous models for galactic global star formation from gas 
in galaxies influenced both by galaxy interactions and tidal fields
of their host larger halos (Bekki \& Chiba 2005) and 
chemical evolution and  supernova feedback effects (Bekki \& Shioya 1999),
we model physical processes of star formation
and chemical evolution   as follows: 
The gas mass fraction ($f_{\rm g}$) is assumed
to be a free parameter ranging from 0.1 to 0.5.
An isothermal equation of state is used for the gas with
temperatures of $10^4$ K for models with $M_{\rm d}=6 \times
10^{10} {\rm M}_{\odot}$ and  the initial temperature ($T_{\rm iso}$) of disk
gas
is assumed to depend on $M_{\rm gal}$.
The gas disk of a spiral is represented by SPH particles and has
an exponential radial density profile that is exactly the same as
the stellar disk has.

In order to construct as realistic a gas disk model as possible,
we consider the radial dependence of the gas mass fraction 
$F_{\rm g}(r)$ in the initial  disk.
We expect that the inner gas mass fraction in the disk is
smaller than the outer value owing to more rapid consumption of gas
in the inner regions with higher gas density.
We therefore adopt the following rule:
\begin{equation}
F_{\rm g}(r) \propto t_{sf} (r)
\propto  \frac{{\Sigma}_{\rm g}(r)}
{\dot{\Sigma}_{\rm g}(r)}
 \propto {{\Sigma}_{\rm g}}^{\alpha} (r),
\end{equation}
where $r$,  $t_{sf}$, ${\Sigma}_{\rm g}$, 
$\dot{\Sigma}_{\rm g}$, and ${\alpha}$ are 
the distance from the center of the  disk, the gas consumption
time scale, the initial gas density,
the gas consumption rate, and the parameter controlling the radial dependence.
Since we adopt the Schmidt law (Schmidt 1959) with  exponent of  1.5  
for star formation (described below),
a reasonable value of $\alpha$ is $-0.5$.

According to the value of $F_{\rm g}(r)$ at each radius
derived from the above equation (4),
we determine a reasonable number of stellar and gaseous particles 
at each radius and thereby allocate these particles to each radial bin. 
By assuming that the disk is composed only of gas initially,
we determine $F_{\rm g}(r)$ through equation (4)
and derive a reasonable radial distribution  of gas and stars.
If we adopt models with constant $F_{\rm g}(r)$, then
a strong initial starburst can occur in the central region of a gas-rich spiral.
We therefore consider that this starburst in an isolated spiral is not reasonable
and realistic. Furthermore the adopted models with 
radial dependence in $F_{\rm g}$ (with no initial starbursts) enable
us to clearly understand how galaxy interactions trigger starbursts within 
the bulges of spirals.

Star formation
is modeled by converting  the collisional
gas particles
into  collisionless new stellar particles according to the algorithm
of star formation  described below.
We adopt the Schmidt law 
with exponent $\gamma$ = 1.5 (1.0  $ < $  $\gamma$
$ < $ 2.0, Kennicutt 1998) as the controlling
parameter of the rate of star formation.
The amount of gas 
consumed by star formation for each gas particle
in each time step 
is given as:
\begin{equation}
\dot{{\rho}_{\rm g}} \propto  
{\rho_{\rm g}}^{\gamma},
\end{equation}
where $\rho_{\rm g}$ 
is the gas density around each gas particle. 
These stars formed from gas are called ``new stars'' (or ``young stars'')
whereas stars initially within a disk  are called ``old stars''
throughout this paper.

Chemical enrichment through star formation and supernova feedback
during the evolution of the disk 
is assumed to proceed both locally and instantaneously in the present study.
We assign the metallicity of the original
gas particle to  the new stellar particle and increase 
the metallicity of each neighboring gas particle 
with the total number of neighbor gas particles equal to  $N_{\rm gas}$,
according to the following 
equation that governs chemical enrichment:
\begin{equation}
\Delta M_{\rm Z} = \{ Z_{i}R_{\rm met}m_{\rm s}+(1.0-R_{\rm met})
(1.0-Z_{i})m_{\rm s}y_{\rm met} \}/N_{\rm gas} 
\end{equation}
where $\Delta M_{\rm Z}$ represents the increase in 
metallicity for each gas particle. $ Z_{i}$, $R_{\rm met}$, $m_{\rm s}$,
and $y_{\rm met}$  in the above equation represent
the metallicity of the new stellar particle (or that of original gas particle),
the fraction of gas returned to the interstellar medium,  the
mass of the new star, and the chemical yield, respectively.

Since our main focus is morphological transformation of luminous spirals
(like the Galaxy) into S0s, we mainly investigate chemical evolution
of the MW models. For models with $M_{\rm d}=6 \times 10^{10} {\rm M}_{\odot}$,
the values of $R_{\rm met}$,  $y_{\rm met}$, and initial metallicity  are set to
be 0.3 and 0.01, and 0.01 (i.e., [Fe/H]=$-0.3$), respectively. For comparison,
we also investigate models with the initial metallicities of 0.02 (the solar
abundance). We adopt the same model for supernova feedback effects (on ISM of
galaxies) as used in our previous chemodynamical simulations (Bekki \& Shioya 1999).
In the present study, all of the energy from a supernova ($10^{51}$ erg per 
a supernova) is converted into kinematic energy of gas around the supernova.

\subsection{Orbits of spirals within groups}

We investigate the dynamical and chemical evolution and star formation
history of a  spiral galaxy in a group for  a given model: only one galaxy is modeled
fully self-consistently by N-body particles and other galaxies orbiting the group are
represented by point-mass particles with the spherical
Plummer softening lengths.
The orbit of our model spiral  is assumed to be
influenced both  by the gravitational potential resulting from the dark halo
component of the group and by the group member galaxies.
The adopted group potential
is spherically symmetric (not triaxial),
and thus the initial orbital plane of the  spiral 
is  set to be  the $x$-$y$ plane (=disk plane
of the spiral)  in all
our models. Owing to dynamical friction of the  spiral against the dark matter
halo of the  group,  the spiral can sink into the central region of the group
if the total mass of the spiral is large enough.

The center of the  group  is always set to be ($x$,$y$,$z$) =
(0,0,0) whereas the initial position of the spiral is set to be ($x$,$y$,$z$)
= ($R_{\rm ini}$, 0, 0). 
The initial velocity of the spiral
($v_{\rm x}$,$v_{\rm y}$,$v_{\rm z}$) is set to be (0, $f_{\rm v} V_{\rm
c}$, 0), where $f_{\rm v}$ and $V_{\rm c}$ are the parameters controlling
the orbital eccentricity (i.e, the larger $f_{\rm v}$ is, the more circular
the orbit becomes) and the circular velocity of the group at
$R$ = $R_{\rm ini}$, respectively. For these models, 
$r_{\rm s} \le R_{\rm ini} \le r_{\rm vir}$
and $0.25 \le f_{\rm v} \le 1.0$, and thus  spirals can initially orbit
the centers of their host groups. 
These models with $R_{\rm ini} \le r_{\rm vir}$ are referred to as
``normal models''.

We also investigate ``infall models''
in which spirals are initially located outside the virial radius of their host
groups.  For these models, $R_{\rm ini}$ is larger than $r_{\rm vir}$ 
(and $y$ is not 0) and $v_{\rm x}=f_{\rm v}V_{\rm c}$
(i.e., $v_{\rm x}$ is not 0 and instead $v_{\rm y}=0$).
In infall models,  spirals move initially parallel  to  the $x$-axis
toward the negative  $x$ direction and the initial value of $y$ corresponds
to the impact parameter ($b_{\rm y}$)  of the orbit.
For all infall models, $R_{\rm ini}$ is set to be $7r_{\rm s}$ and
the impact parameter  $b_{\rm y}$ and $f_{\rm v}$
are the two key parameter that determine the orbits within groups.

\begin{figure*}
\psfig{file=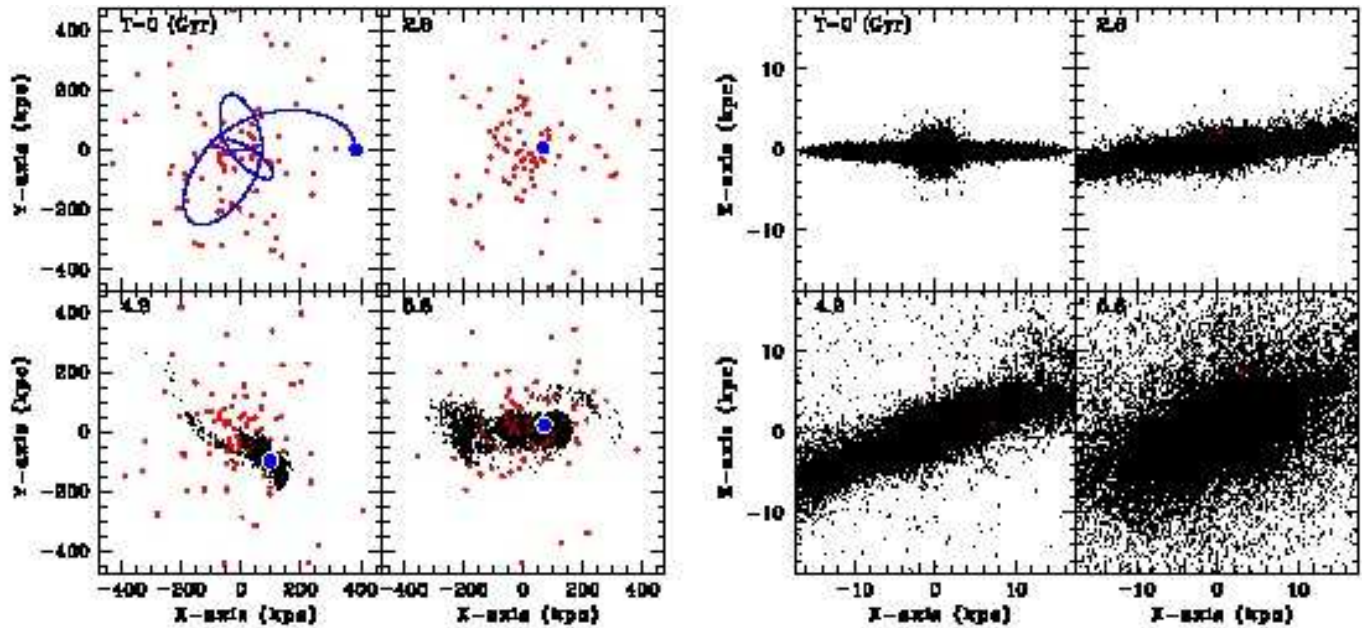,width=18.5cm}
\caption{
The time evolution of  the stellar  distribution of a spiral galaxy projected onto
the $x$-$y$ plane (left four panels) and onto the $x$-$z$ plane (right four panels)
at different times ($T$) for the fiducial model. The time $T$ (in units of Gyr)
is shown in the upper left corner of each panel.  
The solid {\it blue} line in the upper left panel on the left-hand side
represents the orbit of the spiral for the last 5.6 Gyr.
The red particles represent group member galaxies. 
The  upper four
panels on the left and right 
describe 
the time evolution of stellar distributions in smaller- and larger-scales, 
respectively. Stellar particles are not seen in the upper two of the left
four panels, because tidal stripping can not happen before $T=2.8$ Gyr.
In order to clearly show the formation of intragroup stars,
only stars that are stripped from the spiral 
(i.e., those located at $R>2R_{\rm d}$, where $R$ is the distance from the
center of the spiral) are shown in the left four panels. 
}
\label{Figure. 1}
\end{figure*}

\subsection{Choice  of parameters}
\subsubsection{Fiducial model}

We run models with different $M_{\rm d}$, $R_{\rm d}$, $f_{\rm b}$,
$f_{\rm g}$, $M_{\rm gr}$, $R_{\rm ini}$, and $f_{\rm v}$ in order
to investigate the formation processes of S0s and their dependences on model
parameters.  The range of each model parameter is
shown in the Table 1.
We mainly show the results of the ``fiducial model''
in which 
$M_{\rm d}$ = 6.0 $\times$ $10^{10}$ ${\rm M}_{\odot}$
$R_{\rm d}$ = 17.5 kpc (i.e., $C_{\rm d}=3.5$ kpc),
$f_{\rm b}=0.17$ (i.e., MW model),
$f_{\rm g}=0.1$,
$M_{\rm gr}=2 \times 10^{13} {\rm M}_{\odot}$,
$R_{\rm ini}=4r_{\rm s}$,
and $f_{\rm v}=0.5$ (i.e., ``normal model'').
This is mainly because this model shows the typical behavior of S0 formation
for spirals in groups. We run a model for $\sim 6$ Gyr in order
to investigate whether the spiral can be transformed into an S0 in the model.

\subsubsection{Multiple gravitational softening lengths}

Since we need to run $\sim 100$ models for allocated computational time for
the GRAPE system composed of multiple GRAPE 7, we need to  use a limited 
number of particles in each simulation.  We run a model which
requires $\sim 100$ CPU hours of the adopted GRAPE 7 system 
for at least $\sim 6$ Gyr evolution of a spiral in a group.
If we use the same particle mass both for stellar particles in a spiral
and for dark matter particles of its host group, then
we need at least $\sim 10^8$ particles for the dark matter.
Numerically, it is very costly for the present work to
run models with such a large particle number, given
the limited CPU  time. We thus  use a large mass  of
$\sim 10^7 - 10^8 {\rm M}_{\odot}$ for the dark matter particles
of a group to dramatically reduce the total particle number of the group.

Owing to this large mass,  dynamical interaction between group dark matter particles
and a disk galaxy can cause unrealistic dynamical heating of the stellar disk
(and possible transformation from spirals into S0s),
which should be avoided in the present study.
We therefore adopt a large  gravitational softening length ($\epsilon$)
for the dark matter particles of a group to significantly suppress
unrealistic tidal heating of a spiral by  the dark matter particles: we use
different softening lengths for the different components in the simulation
(e.g., dark matter halo and disk of a spiral and dark matter halo of its host group).
We confirm that galaxies with large pericenter distances of their
orbits with respect to their host group's center  (thus with no/little tidal galaxy
interaction) 
can not be transformed into S0s with thick disks: 
artificial tidal heating by group dark matter particles
is well suppressed.

The gravitational softening lengths for the dark matter particles of a spiral,
the stellar particles,  and the dark matter particles of its host group
are denoted as ${\epsilon}_{\rm d,d}$, 
${\epsilon}_{\rm d,s}$, 
and ${\epsilon}_{\rm gr}$, respectively.
{We determine $\epsilon$ for each component based on the mean particle-particle
separation within  the half-number radius of  
the spatial distribution of particle for each component.}
 Furthermore,  when two different components interact gravitationally,
the mean softening length for the two components
is applied for the gravitational calculation.
For example, $\epsilon = ({\epsilon}_{\rm d,s}+{\epsilon}_{\rm gr})/2$
is used for gravitational interaction between stellar particles in a disk
and the dark matter particles of the host group.
The gravitational softening lengths for
particles of  the bulge, the gaseous disk, and new stars in a spiral
are the same as ${\epsilon}_{\rm d,s}$, and ${\epsilon}$ for 
group member galaxies (represented by point-mass particles)
is assumed to be the same as  ${\epsilon}_{\rm gr}$.

The total number of particles 
used  in the fiducial model
is 216,700 and 100,000 for a spiral and its host group, respectively,
and ${\epsilon}_{\rm d,d}$,  ${\epsilon}_{\rm d,s}$ and 
${\epsilon}_{\rm gr}$ are set to be 1.79\,kpc 0.25\,kpc,
and 10.78\,kpc, respectively.
As described above, these softening lengths depend on
$R_{\rm d}$ (i.e., smaller for less massive disks).
The total number of dark matter particles in a group
is linearly proportional to $M_{\rm gr}$: it is 500,000 for
the models with $M_{\rm gr}=10^{14} {\rm M}_{\odot}$. 
We consider that as long as we investigate galaxies
with $M_{\rm gal}$ ranging from $\sim 10^{10} {\rm M}_{\odot}$
to $\sim 10^{12} {\rm M}_{\odot}$,
dynamical friction of the galaxies against the dark matter halos
of their host groups can be properly investigated for such
particle numbers used for groups.
Also
the total number of bulge particles is also linearly proportional
to $M_{\rm b}$ ($=f_{\rm b}M_{\rm d}$) so that models with larger
$f_{\rm b}$ have larger bulge particle numbers.

\subsubsection{Main points of analysis}

We consider that if a  disk galaxy has no spiral arms (by naked-eye inspection)
at the final time step of each simulation, then the simulated galaxy 
can be classified as an S0 galaxy. This is consistent with
the canonical classification method that has been used in observational studies
(e.g., Sandage 1961).
We also use the smoothed two-dimensional (2D) stellar distributions
of simulated galaxies  to demonstrate
more clearly that the simulated S0s have no spiral arms if they are projected onto the
sky. The method to derived the smoothed 2D distributions is given in the Appendix A.
We investigate star formation histories,
the mass fractions of stripped gas and stars, maximum rotational speeds 
of stars ($V_{\rm m}$),   velocity dispersions of gas and stars ($\sigma$),
and evolution of the chemical abundances of the disks and nuclear regions.
We also investigate the spatial distributions of the stripped gas and stars, which
may well be identified as intragroup gas and stars.

Our particular interest is in how spirals with different Hubble-types
(i.e., different $f_{\rm b}$) evolve as a result of galaxy interactions in groups.
The Hubble-types Sd, Sc, Sbc, Sb, and Sa (from late- to early-types in order)
have $f_{\rm b}=0.02$, 0.10, 0.19, 0.32, and 0.70,
respectively,  in the present model
(or $B/T=0.02$, 0.09, 0.16, 0.2, and 0.41, respectively).
For each model, we investigate the increase of the bulge mass 
($\Delta M_{\rm b}$) in a spiral due to 
galaxy interactions
by estimating the total mass of new stars formed from gas within
$R\le 0.2R_{\rm d}$. The growth rate of a spiral  bulge is measured  by
$\Delta M_{\rm b}/M_{\rm b}(0)$, where $M_{\rm b}(0)$ is the initial mass
of the bulge. 
 We do not adopt the same bulge-disk decomposition technique as used for
estimating $B/T$ from the radial light profiles of S0s in observations
(e.g., Laurikainen et al. 2009).
This is mainly because we can not calculate the spectral energy distribution
at each local region (for estimating the radial light profiles)
owing to the fact that the present  numerical code
has not been yet combined with the stellar population synthesis codes.
In our future papers, we will combine our new chemodynamical code
with a stellar population synthesis one and thus be able to derive $B/T$ 
in the exactly the same manner as observations: we will discuss this important
issue extensively in a forthcoming paper.
In the following, $T$ in a simulation
represents the time that has elapsed since the simulation
started.

\begin{figure}
\psfig{file=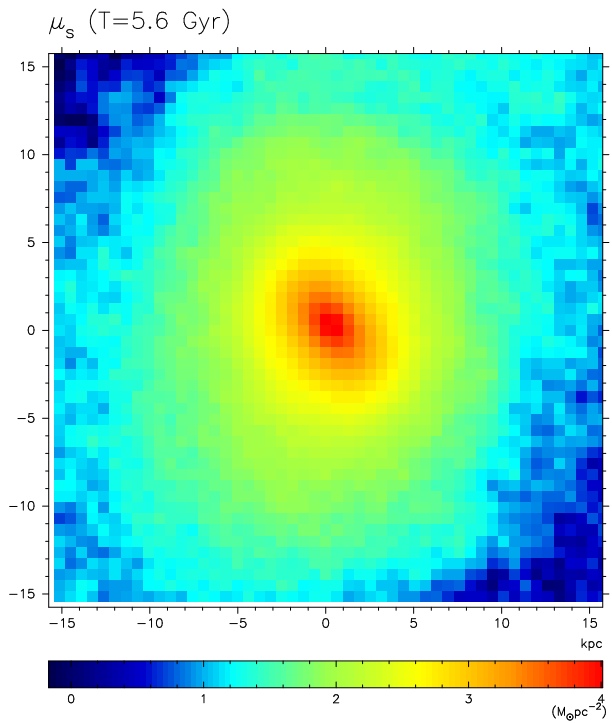,width=8.0cm}
\caption{
The final smoothed 2D density distribution of old stars
projected onto the $x$-$y$ plane in the fiducial model.
This logarithmic density map (${\mu}_{\rm s}$) clearly demonstrates
that the simulated galaxy does not have spirals and thus can be identified
as an S0.
}
\label{Figure. 2}
\end{figure}

\begin{figure}
\psfig{file=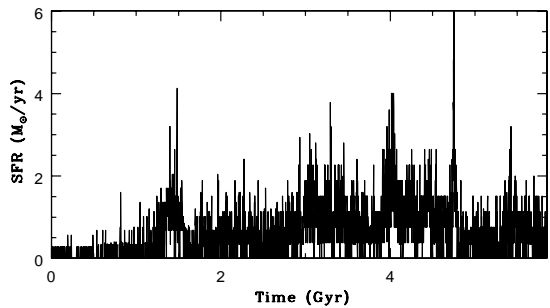,width=8.0cm}
\caption{
The time evolution of the star formation rate in the  spiral for the fiducial model.
}
\label{Figure. 3}
\end{figure}

\begin{figure}
\psfig{file=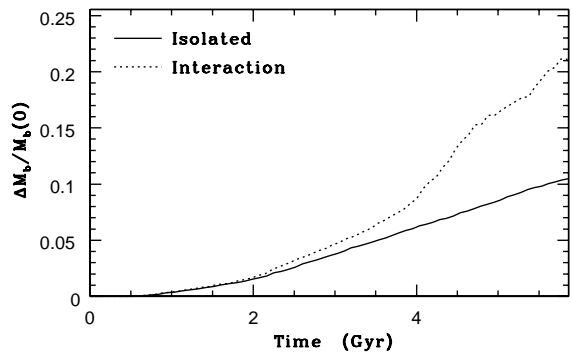,width=8.0cm}
\caption{
The time evolution of the bulge growth rate ($\Delta M_{\rm b}/M_{\rm b}(0)$)
for the fiducial model (dotted) and the isolated model (solid) in which neither
group tidal field nor galaxy interactions are included.
}
\label{Figure. 4}
\end{figure}

\begin{figure}
\psfig{file=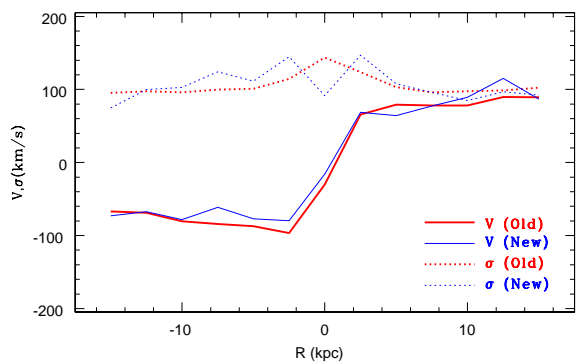,width=8.0cm}
\caption{
Radial profiles of rotational velocities ($V$, solid)
 and velocity dispersions ($\sigma$, dotted)
for old stars (red) and new stars (blue) in the simulated
S0 at $T=5.6$ Gyr for the fiducial model.
Here $V$ and $\sigma$ are derived from line-of-sight velocities 
along the major axis of
the simulated S0 projected onto the $x$-$z$ plane (i.e., edge-on). 
For this estimation,
the $y$-component of a velocity for each individual particle is used.
}
\label{Figure. 5}
\end{figure}

\begin{figure}
\psfig{file=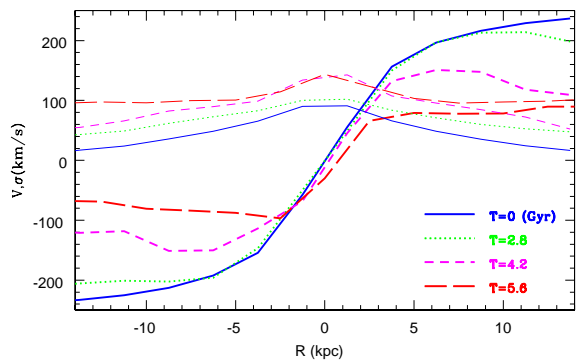,width=8.0cm}
\caption{
Radial profiles of $V$ (thick) and $\sigma$ (thin)
at $T=0$ Gyr (blue solid), 2.8 Gyr (green dotted), 
4.2 Gyr (magenta short-dashed), and 5.6 Gyr (red long-dashed)
for the fiducial model. 
}
\label{Figure. 6}
\end{figure}

\begin{figure}
\psfig{file=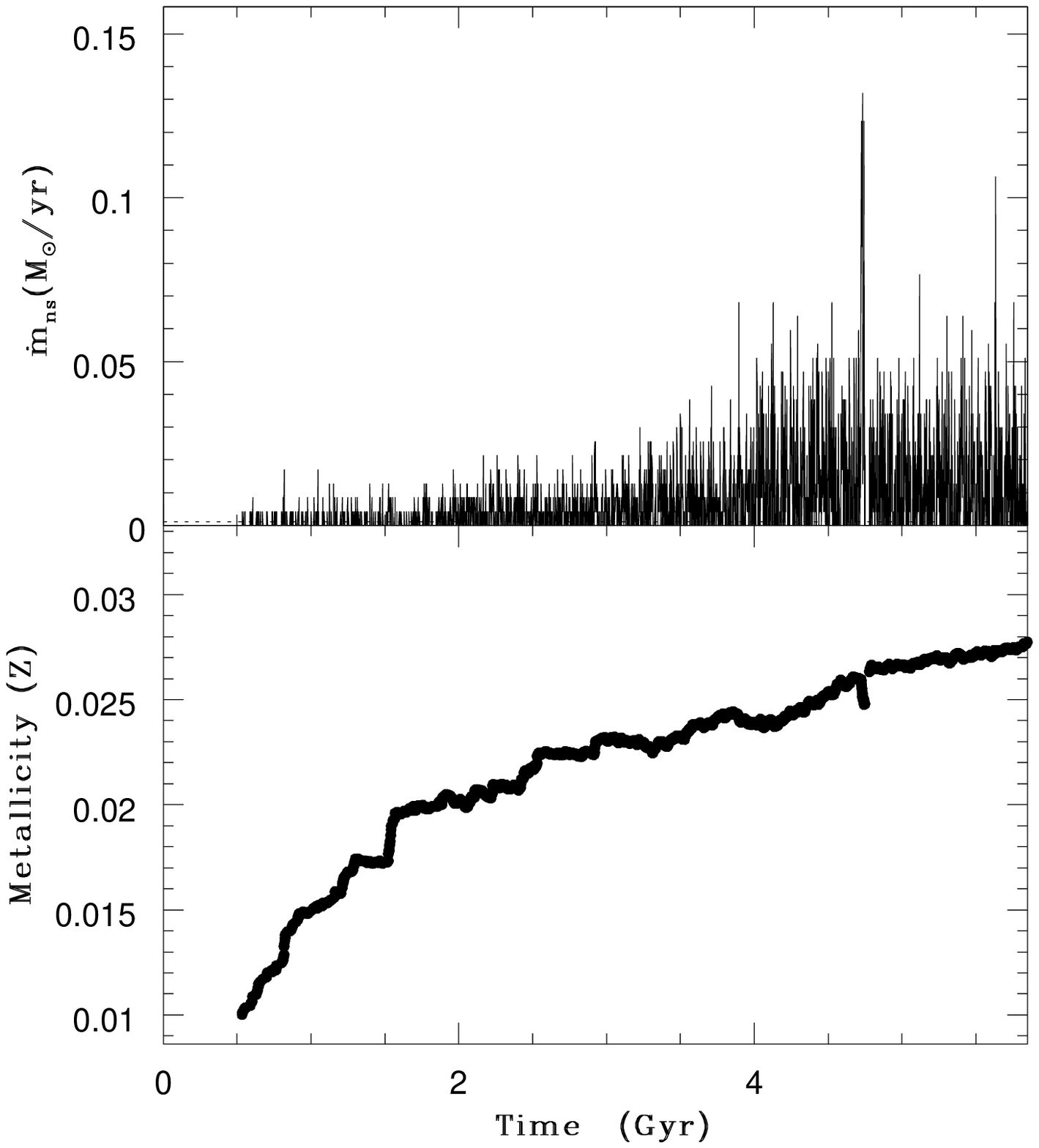,width=8.0cm}
\caption{
The time evolution of the formation rate of new stars within
the central $0.2R_{\rm d}$ (i.e., bulge region) of the spiral (upper)
and the mean metallicity
for the new stars there (lower) for the fiducial model.
}
\label{Figure. 7}
\end{figure}

\section{Results}

\subsection{Fiducial model}

Figure 1 summarizes (i)\,the time evolution
of the orbit and stellar distribution of a  spiral 
in a group, and (ii)\,that of the distribution of group member galaxies, 
for the fiducial model.
As the spiral approaches the pericenter of its orbit for the first time,
the tidal field of the group starts to dynamically heat up 
the stellar disk 
to some extent. The tidal field of the group, however, can only 
influence the outer part of the stellar disk of the spiral.
In the early evolution of the spiral within the group ($T<2.8$ Gyr),
it only encounters low-mass dwarf galaxies so that galaxy
interactions do not dramatically influence the evolution. 
However, as the spiral slowly sinks into the central region 
due to dynamical friction,
it more frequently interacts with group member galaxies due
to the higher number density of the galaxies their.
This sinking of the spiral can not be seen in low-mass disk galaxies
owing to much less effective dynamical friction for the spirals
in the group.
After $T \sim 4$ Gyr,  the spiral
experiences  a number of tidal encounters with galaxies which have comparable
masses (to the spiral) so that the stellar disk is strongly heated up
to form a thick disk  with a vertical velocity dispersion of 
$\sim 60$ km s$^{-1}$ at $R\sim 10$ kpc.

Spiral arms within the disk can gradually disappear
due to the strong tidal heating during the 
interactions with group member galaxies. 
About 24\% (11\%) of the old stars initially in the disk
can be finally located at $R \le 2R_{\rm d}$ and $|z| \ge 0.1 R_{\rm d}$
($R \le 2R_{\rm d}$ and $|z| \ge 0.2 R_{\rm d}$).
This stellar component can therefore be regarded as a stellar halo around the S0
at $T=5.6$ Gyr.
The edge-on view of the 
disk galaxy 
becomes strikingly similar to those of S0s at $T=5.6$ Gyr.
About 11\% of stars can be stripped to be located at $R>2R_{\rm d}$
during the morphological transformation of the spiral into an S0.
These stripped stars may well be identified as intragroup stars drifting freely
in the group. Thus the spiral is transformed into an S0
with a thick disk and a stellar halo as a result of the multiple tidal 
interactions within the group within 5.6 Gyr.

Figure 2 clearly shows that the simulated galaxy  has 
a barred structure yet
no clear spiral-arm structures of old stars in the simulated 2D image at $T=5.6$ Gyr:
this galaxy can be classified as an S0 (or a barred S0).
The new stars have  a very
compact distribution (see Figure A1 in the Appendix A), 
because  star formation in the central
region of the bulge is greatly enhanced in a repetitive manner during
multiple tidal interactions with group member galaxies.
About 60\% of the initial gas can be rapidly consumed by star formation 
so that the final remnant can have a gas mass fraction of only $\sim 0.04$
(i.e., the formation of a gas-poor S0).
The isolated model with no group tide and no galaxy interactions
shows that only 20\% of the initial gas can be converted into new stars
after 5.6\,Gyr of evolution. These results demonstrate that gas-poor S0s
can be formed from multiple tidal interactions in groups.
The final distribution of gas has a weak asymmetric feature (spiral-like morphology),
which is due to the last tidal interaction with a massive group member galaxy.

Figure 3 shows that the spiral experiences a number of moderately strong
``starbursts'' (with the maximum star formation rate of
$\sim 6 {\rm M}_{\odot}$ yr$^{-1}$)  triggered mainly by 
tidal interactions with group member galaxies.
The star formation rates in the intervals between the starbursts are also higher 
than the quiescent level before the spiral is influenced by the group tide ($T<$ 1 Gyr),
which means that interactions with the group member
galaxies can also enhance {\it the average level}  of star formation.
This enhancement is due largely to the significantly increased gas density
in the inner region of the disk.
We have run
a model in which group member galaxies are not included and other model
parameter values are exactly the same as the fiducial model,
and confirmed that such moderately strong repetitive starbursts as shown
in the fiducial model
do not happen in this model (also only 35\% of gas is consumed by
$T=5.6$ Gyr).
This means that multiple tidal encounters are responsible for the significantly enhanced
level of star formation in the spiral.

Figure 4 shows that  (i)\,the bulge of the spiral can increase 
more rapidly in its total mass compared
to the isolated model, 
and (ii)\,the more rapid increase can be seen clearly
at $T>$ 2\,Gyr when the spiral starts undergoing tidal interactions
with group member galaxies.
The final bulge mass is a factor of $\sim 20$\% larger
than its original value in the fiducial model.
This  rapid increase in the bulge mass
results from the fact that a large amount of the ISM can be transferred
to the bulge region in a repetitive manner during multiple tidal
interactions with group member galaxies. The bulge growth
in the isolated model is due to the inner transfer of gas
by dynamical action of a stellar bar in the disk
(i.e., due to ``secular evolution'').

Figures 5 and 6 summarize the final stellar kinematics of the simulated S0
and the time evolution of the kinematics during the morphological transformation.
The kinematical properties are derived along the major axis of the stellar distribution
of the S0 projected onto the $x$-$z$ plane and the $y$-component velocities of
stars are used for radial profiles of line-of-sight rotational velocities
($V$) and velocity dispersion ($\sigma$). 
The rotational profile of the old stars ($V$) 
at $T=5.6$\,Gyr shows the maximum rotation
($V_{\rm m}$) of 112\,km \,s$^{-1}$, which is significantly 
smaller than the original $V_{\rm m}$ ($=220$\,km\,s$^{-1}$).
This is due largely to dynamical heating of the S0 progenitor spiral by galaxy
interactions with group member galaxies.  
The velocity dispersion profile ($\sigma$) of the old stars is rather flat
at $|R|>4$\,kpc and has a peak (${\sigma}_0$) of 144\,km\,s$^{-1}$
(thus $V_{\rm m}/{\sigma}_0=0.78$).

Although the rotational profile of the new stars appears to be more disturbed,
there is no/little difference in the radial profile of $V$ between old and 
new stars: $V_{\rm m}$ ($=128$\,km\,s$^{-1}$) of the new stars is slightly higher
than that of the old stars. 
The new stars have a central dip in their $\sigma$ profile 
(i.e., ${\sigma}_0=92$\,km\,s$^{-1}$), which is seen in most models in
the present study. This smaller ${\sigma}_0$ results from the fact
that the new stars 
of the bulge are formed from gas after it loses a substantial amount
of its kinetic energy  due to gaseous dissipation (mostly by shock) 
during galaxy interaction. 
Given that the central new stars are young and metal-rich
(as described later),
these results mean that the younger, metal-rich populations in the S0 bulge
has different kinematics from the older, metal-poor populations in the bulge.

The time evolution of the $V$ and $\sigma$ profiles demonstrates that
the spiral with a high rotation amplitude ($V_{\rm m}/{\sigma}_0 \sim 3$)
and a steep radial gradient of $\sigma$ can be transformed into an S0
with a lower rotation amplitude ($V_{\rm m}/{\sigma}_0 \sim 1$)
and a flat radial profile of $\sigma$.
It is confirmed that these kinematical properties of the simulated S0
can be seen in different projections (e.g., 
kinematics derived from $v_{\rm z}$ of stars  in the $x$-$y$ projection).
Thus
these results mean that if S0s are the remnant of violent tidal interactions
in groups, then kinematical properties of S0s can be significantly different
from spirals in the sense that S0s are dynamically ``hotter'' than spirals.

Figure 7 shows that gas fueling to the central region ($R \le 0.2R_{\rm d}=3.5$\,kpc)
becomes more efficient after $T \sim 3.5$\,Gyr so that new stars
in the central regions can be formed efficiently.
The more efficient central star formation after $T\sim3.5$ Gyr is due to
more frequent tidal interactions with group member galaxies
that are as massive as the spiral.  Due to
the infall of metal-poor gas from the outer part of the disk,
the mean metallicity of the new stars can become slightly smaller 
at some time steps (e.g., $T\sim 4.7$\,Gyr).  However, such a 
slight decrease of the mean metallicity is a temporary phenomenon, 
and the metallicity can soon start
to increase owing to chemical enrichment.
Thus new stars in the bulge of the S0
have an age-metallicity relation whereby younger stars are likely
to have higher metallicities.
 
\begin{figure}
\psfig{file=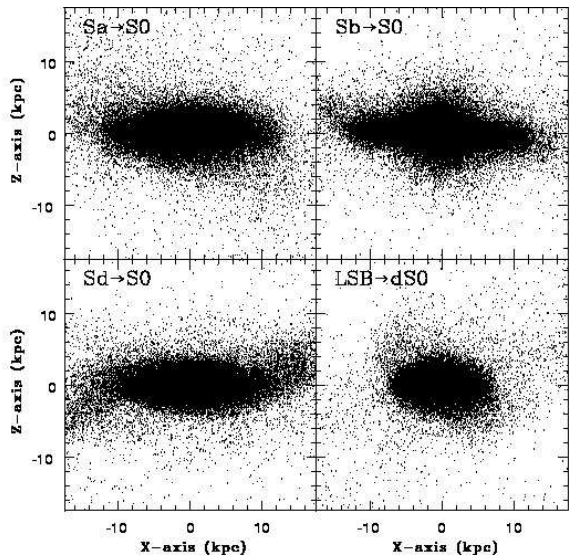,width=8.0cm}
\caption{
The dependence of final morphological properties of the simulated S0s
on the Hubble-types of the S0 progenitor spirals. The results
for the Sa, Sb, Sd, and LSB
models are shown in upper left, upper right,
lower left, and lower right, respectively. 
The models with Sa, Sb, and Sd Hubble-types
have 
$M_{\rm d}=6 \times 10^{10} {\rm M}_{\odot}$
and $R_{\rm d}=17.5$ kpc
whereas the LSB model has 
$M_{\rm d}=6 \times 10^9 {\rm M}_{\odot}$
and $R_{\rm d}=10.0$ kpc.
}
\label{Figure. 8}
\end{figure}

\begin{figure}
\psfig{file=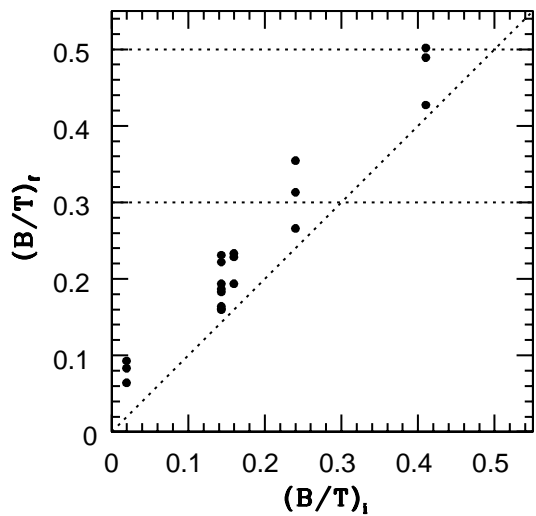,width=8.0cm}
\caption{
The dependence of ${(B/T)}_{\rm f}$ on ${(B/T)}_{\rm i}$
for the simulated S0s in
the representative 20 models with different $f_{\rm b}$, $f_{\rm g}$
and orbits in a group with $M_{\rm gr}=2 \times 10^{13} {\rm M}_{\odot}$. 
The dotted line represents the border line
where ${(B/T)}_{\rm f}={(B/T)}_{\rm i}$
whereas the two horizontal dotted lines show the observed range of $B/T$ that typical S0s
have. 
Here it should be stressed that the simulated $B/T$ are not derived
in the same way as done in recent observations (which uses bulge-disk decomposition
methods). The increase in $B/T$ in the present simulation is due to the mass increase
of new stars in the simulated spiral galaxy.
}
\label{Figure. 9}
\end{figure}

\begin{figure}
\psfig{file=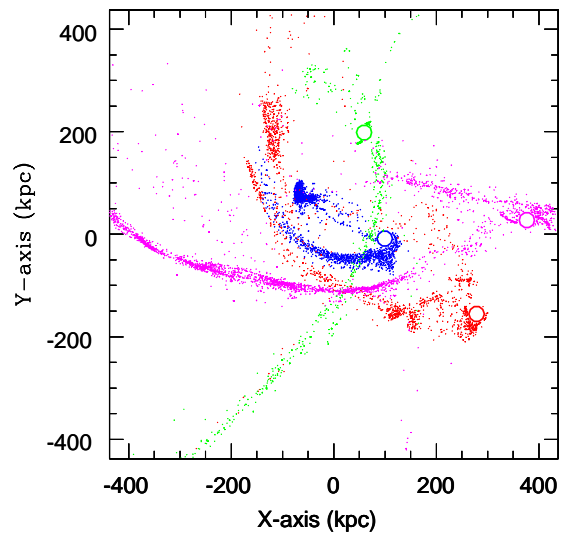,width=8.0cm}
\caption{
The distribution of stripped HI gas projected onto the $x$-$y$ plane
in a group with $M_{\rm gr}=2 \times 10^{13} {\rm M}_{\odot}$
for the low-mass LSB  models
with $M_{\rm d}=6 \times 10^9 {\rm M}_{\odot}$,
$R_{\rm d}=10.0$ kpc,
$f_{\rm b}=0.17$,
$R_{\rm ini}=2r_{\rm s}$,
and $f_{\rm v}=0.25$ (blue),
with $M_{\rm d}=1.2 \times 10^9 {\rm M}_{\odot}$,
$R_{\rm d}=6.6$ kpc,
$f_{\rm b}=0.17$,
$R_{\rm ini}=4r_{\rm s}$,
and $f_{\rm v}=0.25$ (red),
with $M_{\rm d}=1.2 \times 10^9 {\rm M}_{\odot}$,
$R_{\rm d}=6.6$ kpc,
$f_{\rm b}=0.17$,
$R_{\rm ini}=6r_{\rm s}$,
and $f_{\rm v}=0.25$ (green),
and with $M_{\rm d}=1.2 \times 10^9 {\rm M}_{\odot}$,
$R_{\rm d}=6.6$ kpc,
$f_{\rm b}=0.17$,
$R_{\rm ini}=4r_{\rm s}$,
and $f_{\rm v}=0.5$ (magenta).
The gas particles located at $R>2R_{\rm d}$ are regarded as being
stripped from the spiral in each model in this figure.
The big circles represent the final locations of the S0 galaxies 
in the four models.
}
\label{Figure. 10}
\end{figure}

\subsection{Parameter dependences}

Although transformation processes from spirals into S0s
are similar between the fiducial model and other models in which
S0s are formed,
the details of the transformation processes depend on model parameters. 
Also not all of the present models with different
model parameters show such morphological transformation
into S0s.
We illustrate here the derived dependences on the model parameters.

\subsubsection{Orbit}

Spirals in models with
smaller $f_{\rm v}$ (thus smaller pericenter distance $r_{\rm p}$
and larger orbital eccentricity $e_{\rm p}$) 
are more likely to be transformed into S0s
for a given set of other parameters,
because they can be more severely influenced by group tide and 
more frequently interact with group member galaxies in the central regions
of groups.  Also spirals in models with smaller $f_{\rm v}$ can experience
stronger starbursts and thus grow their bulge components 
to a larger extent owing to a larger
amount of gas transferred to the central regions and converted into new
stars there. For example, the model in which  $f_{\rm v}=0.75$ 
and other model parameters are the same as those in the fiducial model,
does not undergo the morphological transformation from a spiral into an S0
seen in the fiducial model. Furthermore, models with smaller
$R_{\rm ini}$ are more likely to be transformed into S0s
for a given set of other parameters.

\subsubsection{Hubble-type}

In the models with $M_{\rm gr}=2\times 10^{13} {\rm M}_{\odot}$ 
and $R_{\rm ini}=4r_{\rm s}$ and $f_{\rm v}=0.25$ and 0.5,
spirals with different $f_{\rm b}$ (thus different Hubble-types)
can be transformed into S0s as a result of the disappearance of 
their initial spiral arms.
However, bigger bulges in the models with larger $f_{\rm b}$ can
severely suppress the gas transfer to the central regions of disks
so that secondary starbursts are not so strong in the models.
This is mainly because (i)\,stellar bars play a role in radial transfer
of gas within disk, 
and (ii)\,the formation of strong bars  during tidal interaction can be suppressed
by the presence of big bulges.
Figure 8 shows the edge-on views of S0s formed from spirals with different initial
$f_{\rm b}$ (i.e., Hubble-types). One of the common features of these simulated S0s
are that they have thick disks and metal-rich  stellar halos.

\subsubsection{Gas mass fraction}

Although the spiral to S0 morphological transformation processes 
do not depend strongly on $f_{\rm g}$,  star formation histories 
are quite different between models with different $f_{\rm g}$ for
a given set of other model parameters.  Due to higher star formation rates 
in the central regions of spirals,
the bulges can grow more rapidly and to a larger extent in models
with larger $f_{\rm g}$. 
For example, the model in which $f_{\rm b}=0.5$ and other model parameters
are the same as those in the fiducial model shows $(B/T)_{\rm f}=0.22$
(in comparison with $(B/T)_{\rm f}=0.18$ for the fiducial model).
These results imply that the bulge growth rates of spirals during
morphological transformation into S0s in groups depend on the 
gas mass fraction of the spirals when they enter into group environments.

\subsubsection{LSB or HSB}

In the present study, 
low-mass galaxies are assumed to have low mean surface stellar densities (LSBs)
so that their evolution in groups can be different from the evolution of  
high-mass galaxies (thus HSB) described above. 
Irrespective of $M_{\rm gr}$,  spirals in models with smaller $M_{\rm gal}$
($\sim 1.2 \times 10^{10} {\rm M}_{\odot}$) are strongly influenced
by both the group tide and by the galaxy interactions with group member galaxies.
As a result of this,  the outer parts of
stellar disks in these less massive spirals  are more severely  disrupted  
and a larger fraction ($\sim 30$\%) of the old stars are tidally stripped to become
intragroup stars. The remnants of this violent tidal interaction appear to
be dwarf S0s (dS0).  Figure 8 shows one example of this dS0 formation
for the model in which $M_{\rm d}=1.2 \times 10^9 {\rm M}_{\odot}$ 
and other model parameters are the same as those in the fiducial model.
Such dS0s can have compact stellar structures in the central regions so that
they can be classified as nucleated dS0s.
Thus the present simulations show that less massive spirals are more likely 
to be transformed into S0s in groups.

\subsubsection{Group mass}

Spirals in  more massive group models with $M_{\rm gr}=5 \times 10^{13} {\rm M}_{\odot}$
are less strongly influenced by group tide and interacting galaxies
in comparison with those in models with
$M_{\rm gr}=2 \times 10^{13} {\rm M}_{\odot}$ for the same $R_{\rm ini}/r_{\rm s}$
($\sim 4$) and $f_{\rm v}$ ($0.25-1.0$). As a result of this,
spirals can continue  their gradual star formation and keep their spiral-arm
structures in the models:
transformation from spirals into S0s due to tidal interactions
is less likely in more massive groups for the above orbits. 
This is true for other massive group models
with $M_{\rm gr}=10^{14} {\rm M}_{\odot}$. 
The main reason for this is that the
relative velocities of two interacting galaxies are higher
in more massive group models
so that galaxy interaction cannot so strongly influence the evolution
of spirals.
It should be stressed, however, that if spirals are initially located in the
central regions ($<2r_{\rm s}$) in these more massive group models, then
spirals can also be rapidly transformed into S0s due to the strong group
tidal field and frequent galaxy interactions there.

\subsubsection{Infall model}

In infall models, it depends on the impact parameter, $b_{\rm y}$, 
and $f_{\rm v}$ whether spirals are strongly influenced by
group tide and galaxy interactions so that their evolution can be significantly
changed in comparison to isolated evolution. 
For models with $M_{\rm gr}=2\times 10^{13} {\rm M}_{\odot}$,
spirals cannot be influenced strongly by tidal effects  in the group as long
as $b_{\rm y}>r_{\rm s}$ (even for $f_{\rm v}=0.2$).
Infalling spirals with $b_{\rm y} \sim 0.5r_{\rm s}$ and $f_{\rm v}=0.2$
can finally have very eccentric orbits within the host group
so that they can be strongly influenced by the group tide and the group
member galaxies when they are in the central region of the group 
(i.e.,  at their pericenter passages).
Due to the longer orbital periods of infalling spirals,
the timescale for them to be transformed into S0s due to galaxy interactions 
is longer.

\subsubsection{Bulge growth rate}

Figure 9 shows the dependences of $(B/T)_{\rm f}$ on $(B/T)_{\rm i}$  
for the 20 simulated S0s
in a group with $M_{\rm gr}=2 \times 10^{13} {\rm M}_{\odot}$
for models with $M_{\rm d}=6 \times 10^{10} {\rm M}_{\odot}$
yet different $f_{\rm b}$,  $f_{\rm g}$, and orbits.
Clearly the $B/T$ can significantly increase in spirals
with different Hubble-types (i.e., $f_{\rm b}$) during their morphological
transformation into S0s due to multiple galaxy interactions 
within the group. However, the final $(B/T)_{\rm f}$ 
for spirals with $(B/T)_{\rm i} <0.2$ cannot be close to that
typically observed ($\approx 0.3-0.5$; Christlein \& Zabludoff 2004), which implies that
typical S0s are not likely to be formed from later-type spirals 
(i.e., later than Sbc). Although early-type spirals can be transformed
into S0s with bigger bulges, the original bulge sizes do not change
during the transformation into S0s.

\subsubsection{Formation of intragroup gas}

Figure 10 summarizes the unique distributions of stripped gas from low-mass galaxies
with $M_{\rm d}$ ranging from $1.2 \times 10^9 {\rm M}_{\odot}$
to  $6 \times 10^9 {\rm M}_{\odot}$
in a group with $M_{\rm gr}=2 \times 10^{13} {\rm M}_{\odot}$.
These stripped gas distributions are well outside their initial disks ($R>2R_{\rm d}$) and
are seen to be drifting freely in the intragroup space so that they would be identified as
intragroup HI gas.  The structure and kinematics 
of the intragroup gas depend strongly on
the orbits of their host spirals.  Apparently isolated giant HI clumps
with no/little optical counter parts can be seen  in some models
(e.g., in the LSB model with $M_{\rm d}=6 \times 10^{9} {\rm M}_{\odot}$,
$C_{\rm d}=6.23$, $f_{\rm b}=0.17$,
$f_{\rm g}=0.1$,  $R_{\rm ini}=2r_{\rm s}$, and $f_{\rm v}=0.25$).
The presence of these HI streams and isolated clouds in groups can be possible
evidence for morphological transformation from spirals into S0s (as discussed later).

\section{Discussion}

\subsection{Formation of S0s with bulge growth}

One of the particularly intriguing results in
recent observations on S0 evolution in groups and clusters of galaxies  is
that the number  evolution of S0s is significantly more dramatic
in groups (or poor clusters) of galaxies 
with $\sigma <$ 750 km s$^{-1}$ at $0.1 < z < 0.8$
(Just et al. 2010).
Other key observational properties of S0s, which should be
explained by any models of S0 formation,
is that the bulge fraction $B/T$  (defined as
the ratio of the bulge to total luminosity of a galaxy)
is significantly larger in S0s than in spirals
(e.g., Dressler 1980; Christlein \& Zabludoff 2004).
These two properties are not explained so simply by the
gradual transformation scenarios of S0 formation via ram pressure
stripping (e.g., Abadi et al. 1999) and removal of galactic halo gas
(e.g., Larson et al. 1980; Bekki et al. 2002; Bekki 2009),
because such scenarios do not predict
the significant increase of the mass concentrations of
spiral galaxies.
Previous numerical simulations of the interactions and merging 
between two galaxies showed that
tidal interaction and merging can transfer gas from the outer parts
of disks into the inner regions to  grow
the bulges (e.g., Noguchi 1988; Bekki 1998) and thus may well
be the physical mechanism for S0 formation.

The present numerical simulations for evolution of spirals 
interacting with numerous
group members have shown that multiple slow tidal encounters
can be responsible for transformation from spirals into S0s with
bulge growth. The growth rates of galactic bulges depend on
the masses, initial $B/T$, and orbits of the original spirals for a 
given group,
and thus not all of spirals can become S0s with big bulge ($B/T>0.5$).
Although spirals with Sd morphology-type can significantly grow
their bulges  owing to the formation of strong bars 
and subsequent gas infall onto the central regions
(resulting from dynamical action of bar on ISM) during tidal interaction,
the final $B/T$ can be still smaller than the typical one for S0s. 
Thus S0s with large $B/T$ ($>0.5$) can hardly be formed from later-type
spirals with Sd, Sc, and Sbc Hubble-types  via tidal interactions in groups.
Other physical mechanisms that increase significantly bulge masses
(e.g., minor and unequal-mass merging)
are necessary for the formation of S0s with bigger bulges.

The present study has shown that less massive spirals are more likely 
to be transformed into S0s, because they are more likely to interact
with galaxies that are larger than themselves in groups and are more
susceptible to group tidal fields.
This suggests that the observed rapid evolution of S0s in groups
(e.g., Wilman et al. 2009; Just et al. 2010) is due to less luminous
galaxies in groups. Also the present study has shown that 
spirals in the inner regions of groups can be more rapidly and dramatically
transformed into S0s, which suggests that number evolution of S0s can be
more clearly seen in the central regions of groups.
Thus it would be important to investigate observationally how
the number fraction of S0s  depends on redshift in groups for
a given mass-range and whether the dependences are different between
the inner and outer regions of groups. 

\subsection{Origin of the Tully-Fisher relation in S0s}

A number of authors have recently discussed  the origin of S0s using the 
Tully-Fisher relation (TFR; Tully-Fisher 1977) of S0s in different environments
(e.g., Hinz et al. 2003; 
Bedregal et al. 2006; Williams et al. 2010).
Hinz et al. (2003) found (i) a larger fraction of stellar mass in S0s than 
in late-type spirals and 
(ii) a larger scatter ($\sigma \sim 1$ mag) 
of  the TFR  in S0s
than in spirals. They therefore suggested
that S0s are likely to be formed from  minor mergers
rather than from simple truncation of star formation and subsequent
disk-fading of late-type spirals.
Bedregal et al. (2006) found that S0s lie systematically 
below the TFR for nearby spirals and have a larger scatter in the TFR
than spirals and thus suggested that simple disk-fading alone cannot explain
the observations so well.
Williams et al. (2010) found that S0s are on average 
fainter in the $K_{\rm S}$ band at a given rotational velocity and suggested
that the observed offset can not be simply  explained  by
disk-fading of spiral galaxies. 

The present simulations have shown that if S0s are formed from spirals
via tidal interaction in groups, then S0s can have smaller maximum
rotational velocity ($V_{\rm m}$)
than their progenitor spirals owing to tidal heating of the disks. 
Although gas can be converted into new stars more efficiently  during
S0 formation via tidal interactions, some fraction ($10-30$\%) of old stars initially
in the disks of spirals can be stripped to become intragroup stars.
Therefore, it is  unlikely that S0s have systematically  larger disk
masses in comparison with spirals. 
Our simulations show that the maximum circular velocities ($V_{\rm c,m}$) of
S0s are very similar to or only slightly lower than
those of their original spirals, though $V_{\rm m}$ of S0s are significantly
lower than spirals: $V_{\rm c, m}=245$ km s$^{-1}$ for the original spiral
and 235 km s$^{-1}$ for the final S0 at $T=5.6$ Gyr (due to mass loss)
in the fiducial model.
Our simulations therefore imply that S0s formed from tidal interactions 
can show  $V_{\rm m}$ similar to or only slightly lower
than  spirals for
a given luminosity a few Gyr after their formation
(when the secondary starburst populations formed during
S0 formation via tidal interaction have faded out). 
Observational results by Williams et al. (2010)
do not show systematically lower $V_{\rm c, m}$ in S0s for 
a given luminosity, though their S0s are not within groups
and their $V_{\rm c, m}$ values are derived from gas (not stars as done in 
the present simulations).

As pointed out by many authors,  there could be a number of viable physical
mechanisms for S0 formation (see \S 1 in this paper).
If S0s are formed from 
disk-fading after truncation of star formation due to removal of
galactic halo gas (e.g., Larson et al. 1980; Bekki et al. 2002), 
then S0s are likely to lie systematically below the TFR of spirals
(e.g., Bedregal et al. 2006). If S0s are formed from spirals  by
tidal interaction with some fractions of original dark matter
and disk stars being stripped, then
S0s can show  slightly lower $V_{\rm c,  m}$ for a given luminosity or mass
owing to the mass loss during  tidal interaction. 
Therefore, the observed large scatter in the TFR of S0s 
(e.g., Hinz et al. 2003;  Williams et al. 2010) appears to be
consistent with S0 formation via different physical mechanisms. 
It is currently impossible to discuss whether the TFR of S0s 
in {\it groups}, where galaxy interactions can be one of the major mechanisms
for S0 formation,  is different to that in the {\it field} and {\it clusters}
due to the lack of observational data sets.
Thus future observations on the TFR in different environments will
enable us to discuss what mechanism(s) dominates S0 formation in the 
field, and in groups and clusters, based on the locations of S0s in the TFR.

\subsection{Formation of intragroup cold gas}

Recent observations have confirmed that nearby clusters have intracluster
planetary nebulae (e.g., Arnaboldi et al. 2003; Arnaboldi 2010 for a recent
review) and globular clusters (e.g. Bassino et al. 2003;
Lee et al. 2010), which implies that
these intracluster stellar objects originate from tidal stripping of
stars and globular clusters from  cluster member galaxies. 
Numerical simulations based on $\Lambda$CDM models
showed that such intracluster stars  can be
formed during the hierarchical assembly of clusters of galaxies
and the simulated radial distributions of the stars are consistent
with the observed ones
(e.g., Murante et al. 2007).
Furthermore,  cosmological N-body simulations with a formation model of 
globular clusters demonstrated that intracluster globular clusters can be formed
from tidal stripping of globular clusters initially in low-mass galaxy-scale
halos (Yahagi \& Bekki 2005; Bekki \& Yahagi 2006).
Although these observational and theoretical studies have discussed
extensively the origin of intracluster {\it stellar} objects in {\it clusters},
they did not discuss the possible existence of intragroup gas.

The present study has demonstrated that if S0s are formed  from 
gas-rich spirals  via  tidal interactions
in groups, then there should be intragroup HI gas that can have unique distributions
like very long tails, broken-rings, and apparently isolated massive clouds.
Therefore it would be reasonable to claim that the presence of such intragroup HI gas
is possible evidence for morphological transformation from spirals into S0s
via tidal interactions.  
Also if S0s are forming in groups at $0<z<0.8$, as suggested by
recent observations (e.g., Just et al. 2010),
then a larger amount of intragroup HI gas should be observed 
in groups at such redshifts.
The present simulations also predict that tidal stripping of gas is more efficient
in the outer parts of disks (where gaseous metallicities are lower
owing to the negative metallicity gradients) for
less massive galaxies  so that the intragroup gas
can be composed mostly of metal-poor gas.
This metal-poor gas would finally mix with the hot gaseous halos of groups
and consequently change the metallicity of the halo gas.

\section{Conclusions}

We have numerically investigated in detail how spirals dynamically interact
with group member galaxies and group tidal fields
and consequently  evolve into S0s 
using chemodynamical simulations of galaxies.
We have analyzed structural, kinematical, and chemical properties
of the simulated S0s in groups for a variety of model parameters
(e.g., $M_{\rm d}$, $f_{\rm b}$, and $M_{\rm gr}$) that can
the control physical properties of galaxies (e.g., Hubble-types) and groups.
We summarize our principle results as
follows.

(1)\,Spirals can be transformed into gas-poor S0s with no remarkable spiral
arms within $\sim 6$\,Gyr evolution of the spirals within groups
with $M_{\rm gr}=2 \times 10^{13} {\rm M}_{\odot}$.
Multiple slow tidal encounters with group member galaxies can dynamically
heat up the disks of spirals so that their initially thin disks can be transformed
into thick ones during morphological transformation from spirals into S0s.
Such tidal interaction can also trigger moderately strong 
(up to $\sim 10 {\rm M}_{\odot}$ yr$^{-1}$ for $f_{\rm g}=0.1$) 
starbursts in the inner regions of 
their bulges so that bulges can grow significantly during S0s formation.
The strong tidal fields of groups can also play a role in morphological
transformation from spirals into S0s for less massive spirals.

(2)\,The details of the processes responsible for the morphological transformation 
of spirals into S0s depend strongly on galaxy masses, the orbits of spirals within groups,
and the total masses of groups ($M_{\rm gr}$).
Spirals initially located in the inner regions of groups can be more rapidly
converted into S0s due to the larger galaxy densities and stronger 
tidal fields of groups there. Less massive spirals are more likely to
be transformed into S0s, because they can interact with galaxies that are 
significantly more massive than themselves and are susceptible to tidal
fields of groups.  

(3)\,Due to significantly enhanced star formation
and efficient chemical enrichment in the inner regions
of spirals during morphological
transformation into S0s,  the bulges of the final S0s
have young and metal-rich stellar populations.  The young, metal-rich
stellar populations are located in the inner regions  
of S0 bulges and have dynamically colder (i.e., smaller ${\sigma}$) than
old stellar populations initially in the bulges. 
The mass fractions of younger and metal-rich populations in S0 bulges
depends strongly on the initial gas mass fractions ($f_{\rm g}$).
Less massive spirals ($M_{\rm d} \approx 10^{10} {\rm M}_{\odot}$) with small bulges
can be transformed into nucleated S0s due to strong nuclear starbursts
occurring during the morphological transformation process.

(4) Stellar disks of S0s formed from tidal interactions in groups
have significantly flatter radial profiles of line-of-sight velocity dispersions
($\sigma$) in comparison with original spirals, because the
outer parts of the original disks 
(where $\sigma$ are initially small) can be dynamically heated up more strongly.
The maximum rotational  velocities ($V_{\rm m}$, which is significantly
lower than 
the maximum circular velocity, $V_{\rm c, m}$) of disks can become
significantly smaller during morphological transformation 
owing to dynamical heating by slow tidal encounters and group tidal fields.
Thus, if S0s are formed from tidal interaction in groups, 
then they can have smaller $V_{\rm m}$ for a given luminosity.

(5) Intragroup cold HI gas can be formed from gas stripped from
spirals being transformed into S0s due to tidal interactions in groups.
Such intragroup gas can have unique spatial distributions (e.g., rings
and very long tails) that reflect the orbits of gas-rich spirals from
which the gas originates from. 
Some intragroup gas clouds  are far away from the galaxies they originated from
so that they  can be identified as apparently isolated massive HI clouds
with no optical counterparts  in groups.
Since most intragroup gas is from the outer parts of disks in less massive
spirals, the metallicity of the gas should be significantly lower.
The presence of intragroup HI gas with unique spatial distributions
is suggested to be possible evidence for morphological transformation
from spirals into S0s due to tidal interaction in groups.

Finally, tidal interactions with group member galaxies  can be  just one of 
a number of possible physical mechanisms for the transformation of spirals into S0s.
S0s with thin disks can be formed from the gradual truncation of star formation
caused by halo gas stripping 
whereas S0s with big bulges ($B/T>0.5$) and with thick disks
can be formed from minor and unequal-mass galaxy merging.
In our forthcoming papers, we discuss the physical properties of S0s formed by
these physical processes that were not explored in the present paper.

\section{Acknowledgment}
We are grateful to the  anonymous referee for valuable comments
which contribute to improve the present paper.
KB  and WJC acknowledge the financial support of the
Australian Research
Council throughout the course of this work. Numerical computations
reported here were carried out both on the GRAPE system at the
University of Western Australia  and on those kindly made available
by the Center for computational astrophysics
(CfCA) of the National Astronomical Observatory of Japan.

\appendix

\section{Derivation of 2D  density fields}

\begin{figure}
\psfig{file=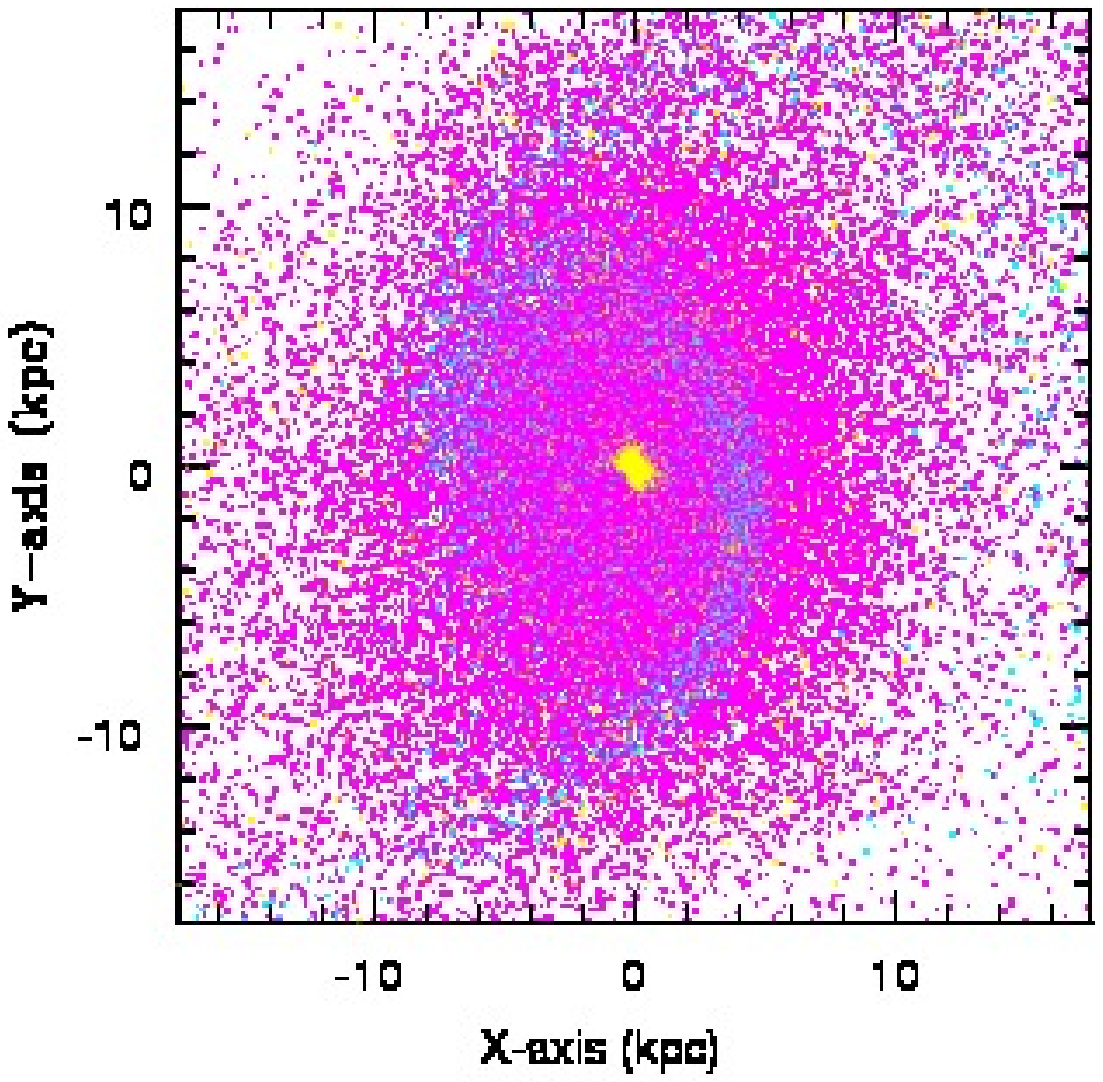,width=8.0cm}
\caption{
The final distributions of old stars (magenta), gas (blue), and new stars (yellow)
projected onto the $x$-$y$ plane in the fiducial model.
}
\label{Figure. 11}
\end{figure}

\begin{figure}
\psfig{file=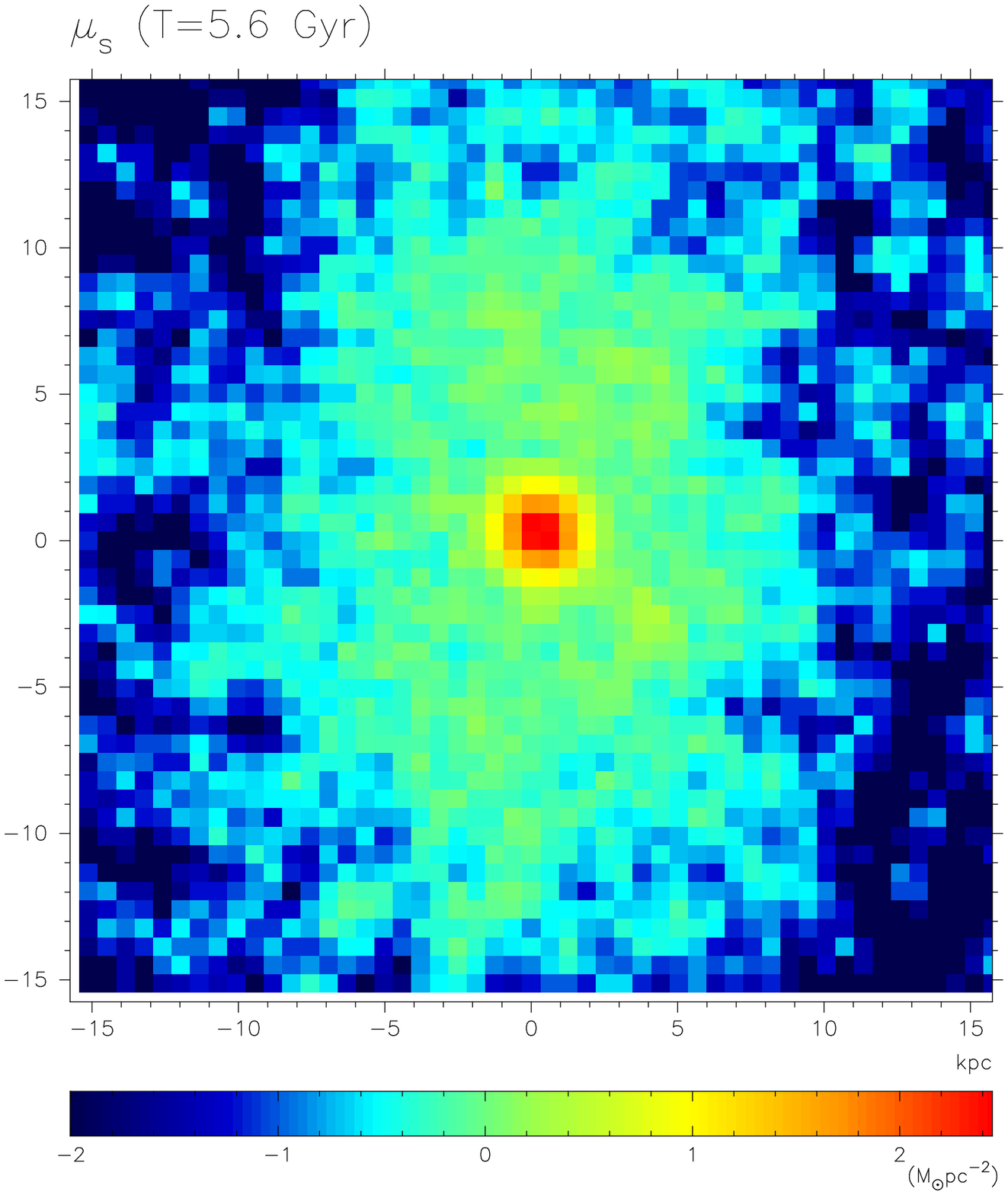,width=8.0cm}
\caption{
The same as Figure 2 but for new stars in the fiducial model.
}
\label{Figure. 12}
\end{figure}

\begin{figure}
\psfig{file=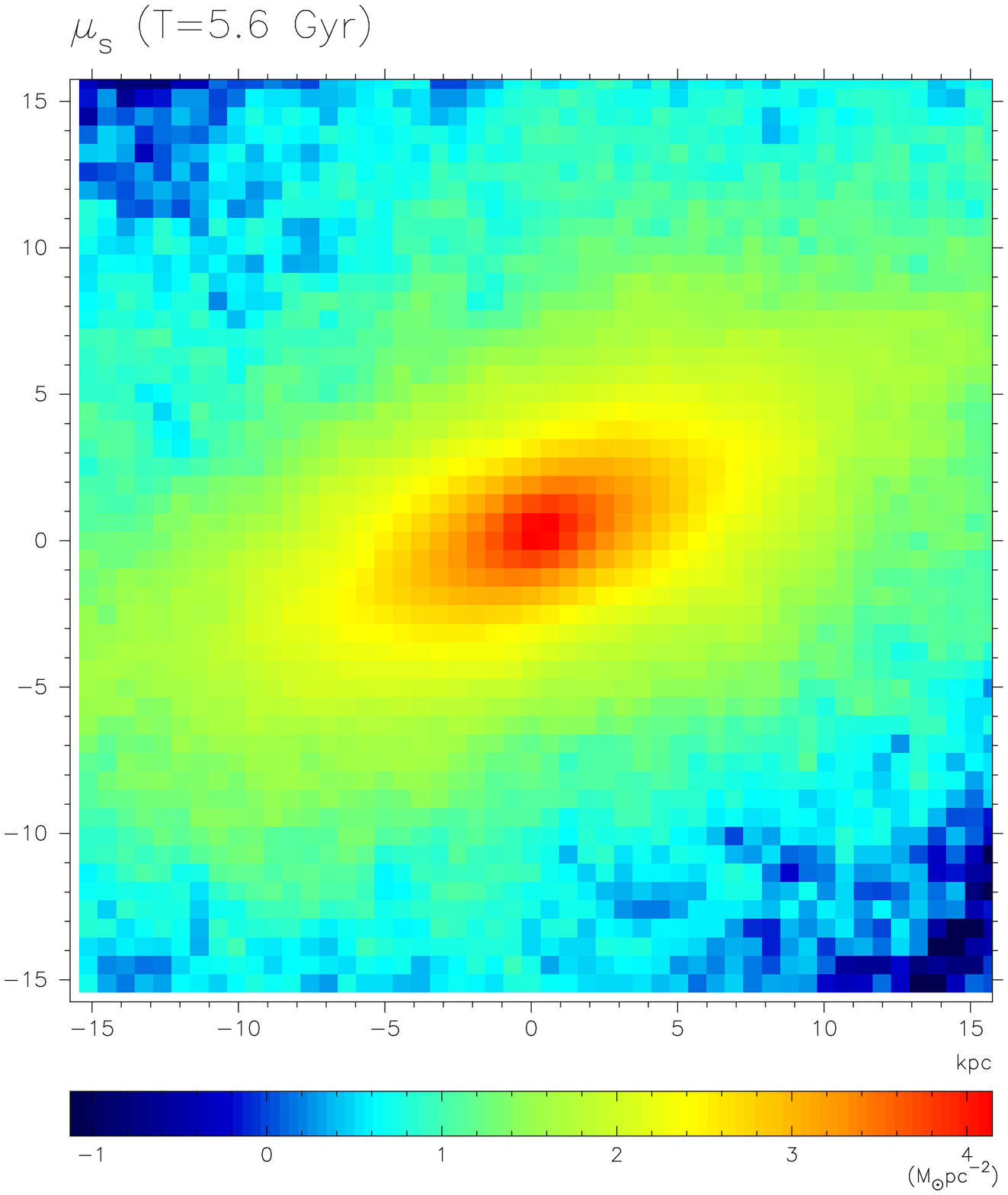,width=8.0cm}
\caption{
The same as Figure 2 but for the $x$-$z$ projection.
}
\label{Figure. 13}
\end{figure}

In order to compare the smoothed 2D density fields of simulated S0s
with the observed images of galaxies (on the sky)  in a more self-consistent
manner, we adopt the smoothing method that is identical  to that used
for analysis of the simulated stellar distributions in the remnants of galaxy mergers
(Bekki \& Peng 2006). We therefore briefly summarized the adopted method
in the present paper.
At the position of each stellar particle,  we apply a local linear smoother
using a Gaussian kernel function with the smoothing length of 0.05 in our
units (corresponding to 0.875 kpc).
We divide the relevant region with the size of $\sim 30$ kpc  (as shown in Fig. 2)
into 50 $\times$ 50 cells
for a model in each projection and estimate the local smoothed stellar density
(${\Sigma}_{\rm s}$). We estimate the 2D logarithmic density map (${\mu}_{\rm s}
={\log}_{10} {\Sigma}_{\rm s}$) for galaxy in  a model in order to mimic the observed image
of the galaxy on the sky.
Total number of cells in each projection
is fixed at 2500 for all  models in the present study, because this number is 
enough to show clearly the presence or absence of the spiral arms in the simulated
disk galaxies.

Figure A1 shows the spatial distributions of old stars, gas, and new stars
at $T=5.6$ Gyr in the fiducial model, which
are used for creating smoothed 2D stellar distributions shown in Figure 2.
Although an arm-like gaseous structure can be discernibly seen in Figure A1,
stellar distributions are overall featureless so that the simulated galaxy
with no spiral arms 
can be classified as an S0.  Figure A2, which shows the smoothed 2D density profile
for new stars,  demonstrates that (i) the young component (formed from
secondary starbursts) of the S0 has a different
spatial distribution in comparison with the old one
and (ii) it also has no clear spiral arm-like structures in the $x$-$y$ projection.
The central young component might well be observed as a ``compact blue bulge'' in
observations. Figure A3 shows the smoothed stellar distribution  of old stars viewed
from edge-on (i.e., $x$-$z$) and confirms that the simulated galaxy appears to be
``lenticular'' like observed S0s. It is clear from this figure that
the S0 has an extended stellar halo that is formed from tidal stripping of stars
during tidal interaction in the group. In order to clearly show how spirals are
transformed into S0s due to tidal interaction,
 we have produced 2D images on  the time evolution of smoothed
2D stellar distributions of galaxies: Figure A4 shows an example for this in the
fiducial model.

\begin{figure}
\psfig{file=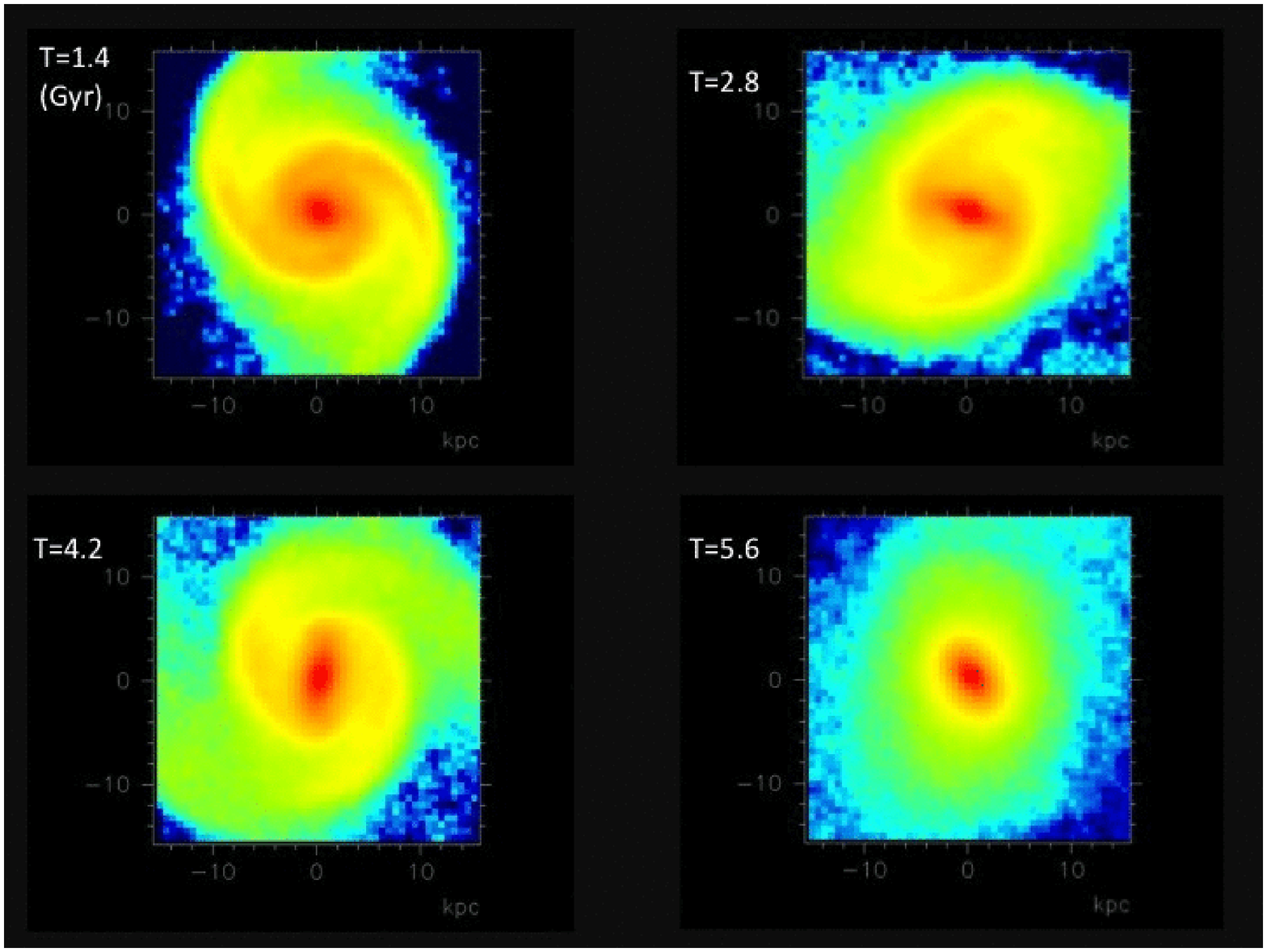,width=8.5cm}
\caption{
The same as Figure 2 but for different four time steps, $T=1.4$ Gyr,
2.8 Gyr, 4.2 Gyr, and 5.6 Gyr.  For convenience,  the color bar shown in Figure 2
to indicate physical mass-densities is not shown in each frame.
}
\label{Figure. 14}
\end{figure}

\section{Bulge-disk decomposition}

Figure B1 shows the
radial mass-density profiles separately for bulge and old stellar disk
at $T=0$ in the original spiral galaxy for the fiducial model. This is simply
a radial density profile and not a luminosity-density profile: the radial 
luminosity-density profile depends strongly on the age and metallicity distributions
of stars in the bulge and disk components. Figure B2 shows the final radial mass-density
profiles separately for the bulge, old stars, and new ones of the simulated
S0 at $T=5.6$ Gyr in the fiducial model.  The overall profile for the total mass-density
is significantly changed between $T=0$ Gyr and $T=5.6$ Gyr owing to tidal stripping of
stars and secondary central starbursts during morphological transformation,
though the central density does not change so much. The stellar
bar formed in the stellar disk during tidal interaction
can interact dynamically  with the original bulge
so that the radial density distributions of the bulge and disk components can
change during the 5.6 Gyr dynamical interaction in the group.
In our future papers,  we try to produce radial light profiles of the simulated S0s
by combining the present chemodynamical code with the stellar population synthesis one
so that we can discuss the origin of S0s in the context of the observed structural
differences of stars between spirals and S0s in a fully self-consistent manner

\begin{figure}
\psfig{file=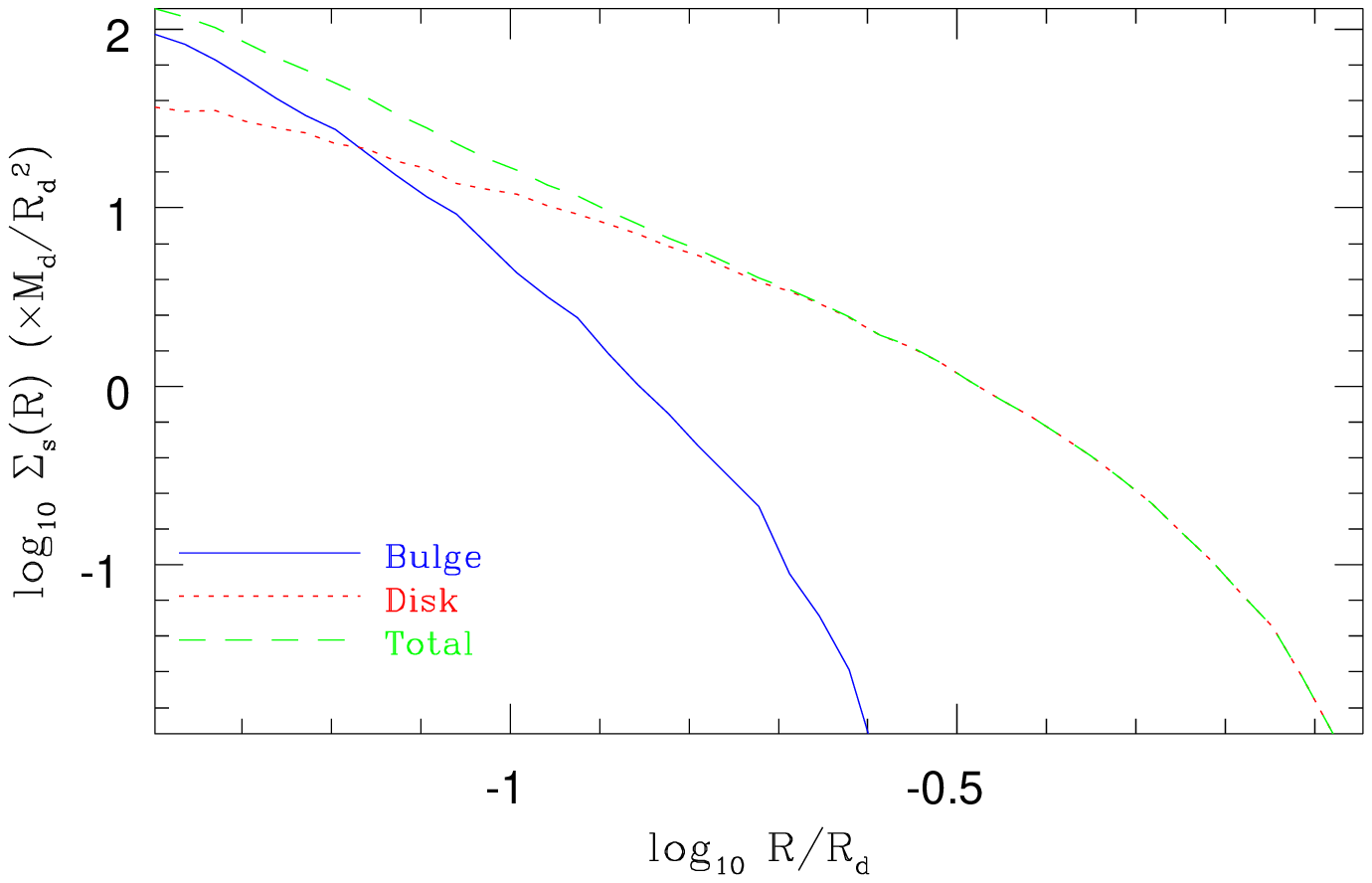,width=8.0cm}
\caption{
The final radial mass-density distributions (${\Sigma}_{\rm s}(R)$)
of bulge stars (solid blue), old disk ones (dotted red), and total (short-dashed green)
projected onto the $x$-$y$ plane at $T=0$ Gyr in the fiducial model.
Here $R_{\rm d}$ describes the initial disk size (=17.5 kpc) in the fiducial model.
}
\label{Figure. 15}
\end{figure}

\begin{figure}
\psfig{file=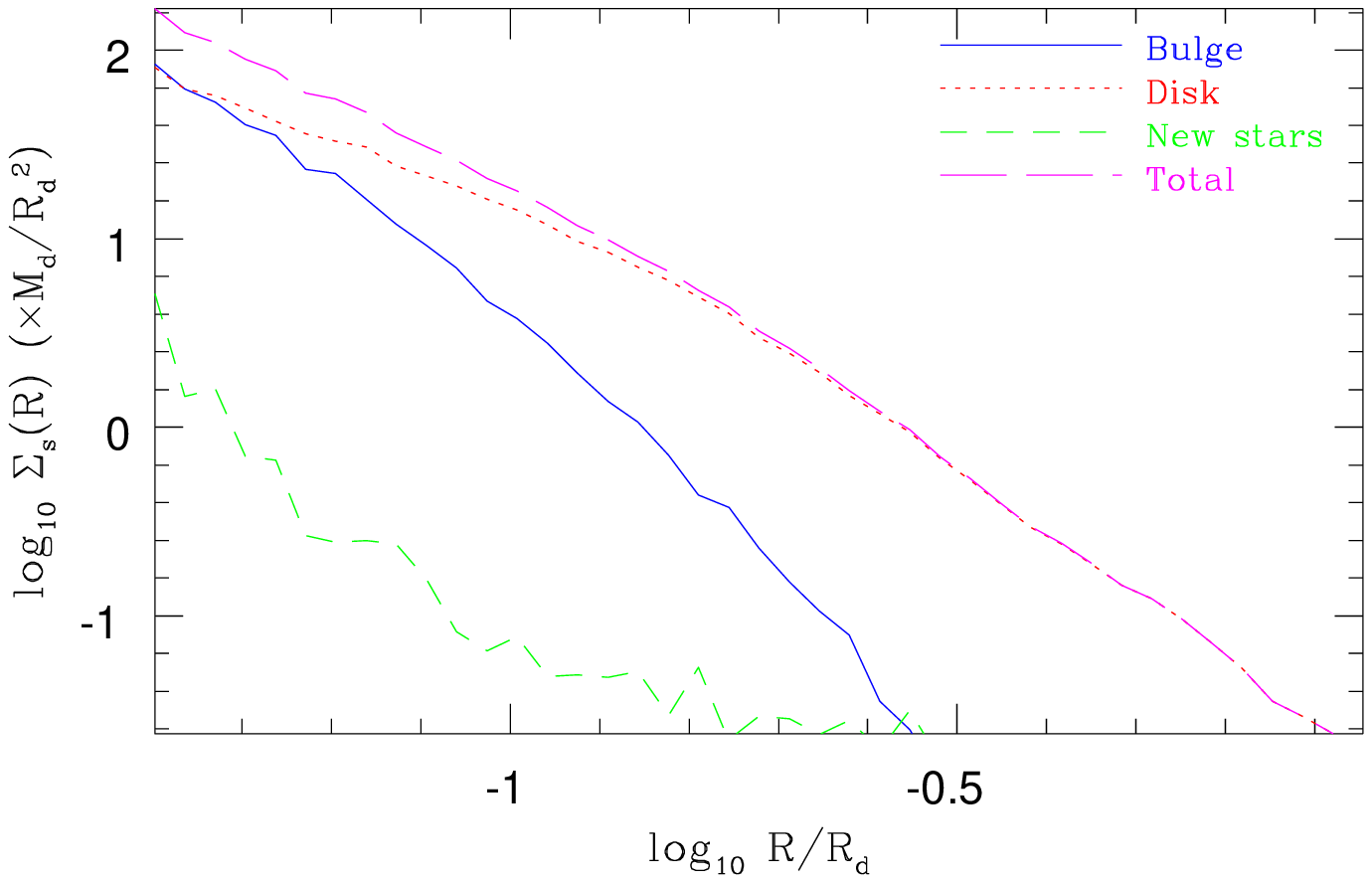,width=8.0cm}
\caption{
The final radial mass-density distributions (${\Sigma}_{\rm s}(R)$)
of bulge stars (solid blue), old disk ones (dotted red), 
new stars (short-dashed green), and total (long-dashed magenta)
projected onto the $x$-$y$ plane at $T=5.6$ Gyr in the fiducial model.
}
\label{Figure. 16}
\end{figure}

\end{document}